\documentclass[twoside,twocolumn,9pt]{article}
\usepackage{extsizes}
\usepackage[super,sort&compress,comma,sort&compress,numbers,super]{natbib} 
\usepackage[version=3]{mhchem}
\usepackage[left=1.5cm, right=1.5cm, top=1.785cm, bottom=2.0cm]{geometry}
\usepackage{balance}
\usepackage{mathptmx}
\usepackage{sectsty}
\usepackage{graphicx} 
\usepackage{lastpage}
\usepackage[format=plain,justification=justified,singlelinecheck=false,font={stretch=1.125,small,sf},labelfont=bf,labelsep=space]{caption}
\usepackage{float}
\usepackage{fancyhdr}
\usepackage{fnpos}
\usepackage[english]{babel}
\addto{\captionsenglish}{%
  
}
\usepackage{array}
\usepackage{droidsans}
\usepackage{charter}
\usepackage[T1]{fontenc}
\usepackage[usenames,dvipsnames]{xcolor}
\usepackage{setspace}
\usepackage[compact]{titlesec}
\usepackage{hyperref}

\usepackage{epstopdf}

\definecolor{cream}{RGB}{222,217,201}

\usepackage{multirow}
\usepackage{xcolor}
\usepackage[normalem]{ulem}

\usepackage{xr}
\makeatletter
\newcommand*{\addFileDependency}[1]{
  \typeout{(#1)}
  \@addtofilelist{#1}
  \IfFileExists{#1}{}{\typeout{No file #1.}}
}
\makeatother

\newcommand*{\myexternaldocument}[1]{%
    \externaldocument{#1}%
    \addFileDependency{#1.tex}%
    \addFileDependency{#1.aux}%
}

\myexternaldocument{SI}
\usepackage[usetitle]{rsc}

\begin{document}

\pagestyle{fancy}
\thispagestyle{plain}
\fancypagestyle{plain}{
\renewcommand{\headrulewidth}{0pt}
}

\makeFNbottom
\makeatletter
\renewcommand\LARGE{\@setfontsize\LARGE{15pt}{17}}
\renewcommand\Large{\@setfontsize\Large{12pt}{14}}
\renewcommand\large{\@setfontsize\large{10pt}{12}}
\renewcommand\footnotesize{\@setfontsize\footnotesize{7pt}{10}}
\makeatother

\renewcommand{\thefootnote}{\fnsymbol{footnote}}
\renewcommand\footnoterule{\vspace*{1pt}%
\color{cream}\hrule width 3.5in height 0.4pt \color{black}\vspace*{5pt}} 
\setcounter{secnumdepth}{5}

\makeatletter 
\renewcommand\@biblabel[1]{#1}            
\renewcommand\@makefntext[1]%
{\noindent\makebox[0pt][r]{\@thefnmark\,}#1}
\makeatother 
\renewcommand{\figurename}{\small{Fig.}~}
\sectionfont{\sffamily\Large}
\subsectionfont{\normalsize}
\subsubsectionfont{\bf}
\setstretch{1.125} 
\setlength{\skip\footins}{0.8cm}
\setlength{\footnotesep}{0.25cm}
\setlength{\jot}{10pt}
\titlespacing*{\section}{0pt}{4pt}{4pt}
\titlespacing*{\subsection}{0pt}{15pt}{1pt}

\fancyfoot{}
\fancyfoot[LO,RE]{\vspace{-7.1pt}\includegraphics[height=9pt]{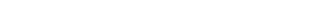}}
\fancyfoot[CO]{\vspace{-7.1pt}\hspace{13.2cm}\includegraphics{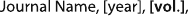}}
\fancyfoot[CE]{\vspace{-7.2pt}\hspace{-14.2cm}\includegraphics{head_foot/RF}}
\fancyfoot[RO]{\footnotesize{\sffamily{1--\pageref{LastPage} ~\textbar  \hspace{2pt}\thepage}}}
\fancyfoot[LE]{\footnotesize{\sffamily{\thepage~\textbar\hspace{3.45cm} 1--\pageref{LastPage}}}}
\fancyhead{}
\renewcommand{\headrulewidth}{0pt} 
\renewcommand{\footrulewidth}{0pt}
\setlength{\arrayrulewidth}{1pt}
\setlength{\columnsep}{6.5mm}
\setlength\bibsep{1pt}

\makeatletter 
\newlength{\figrulesep} 
\setlength{\figrulesep}{0.5\textfloatsep} 

\newcommand{\topfigrule}{\vspace*{-1pt}%
\noindent{\color{cream}\rule[-\figrulesep]{\columnwidth}{1.5pt}} }

\newcommand{\botfigrule}{\vspace*{-2pt}%
\noindent{\color{cream}\rule[\figrulesep]{\columnwidth}{1.5pt}} }

\newcommand{\dblfigrule}{\vspace*{-1pt}%
\noindent{\color{cream}\rule[-\figrulesep]{\textwidth}{1.5pt}} }

\makeatother

\twocolumn[
  \begin{@twocolumnfalse}
{\includegraphics[height=30pt]{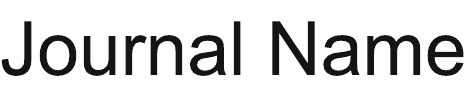}\hfill\raisebox{0pt}[0pt][0pt]{\includegraphics[height=55pt]{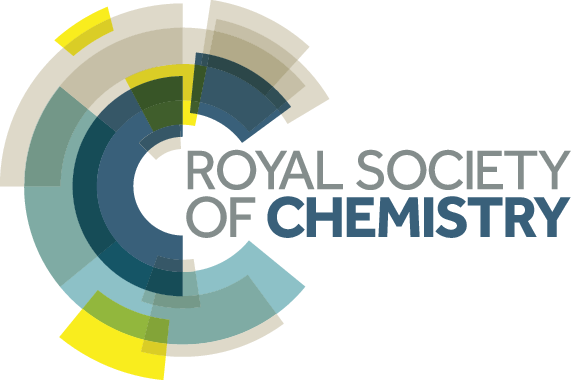}}\\[1ex]
\includegraphics[width=18.5cm]{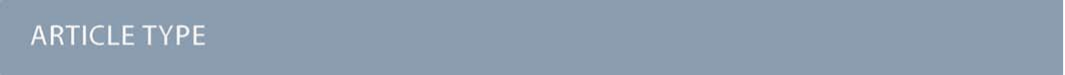}}\par
\vspace{1em}
\sffamily
\begin{tabular}{m{4.5cm} p{13.5cm} }

\includegraphics{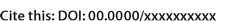} & \noindent\LARGE{\textbf{The effect of mixed termination composition in Sc, Ti, and V-based MXenes$^\dag$}} \\
\vspace{0.3cm} & \vspace{0.3cm} \\

 & \noindent\large{Michal Novotn\'{y}, Karol\'{i}na  Tk\'{a}\v{c}ov\'{a}, and Franti\v{s}ek Karlick\'{y}$^{\ast}$} \\

\includegraphics{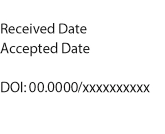} & \noindent\normalsize{In this study, we investigate the effect of mixed surface terminations (F, O, OH) on the properties of M$_{2}$C MXenes (M = Sc, Ti, V). We explore how different compositions and patterns of terminal groups affect the stability and electronic properties of these 2D materials. The bond dissociation energies and cohesion energies show a clear preference for F-terminations in Sc$_{2}$C, while Ti- and V-based MXenes prefer O-terminations. The data indicates that terminal groups on opposite sides of the MXene have little to no influence on each other's electronic structure, allowing for independent chemical environments on each side. Semiconducting forms of studied MXenes (Sc$_{2}$CF$_{2}$, Sc$_{2}$C(OH)$_{2}$ and Ti$_{2}$CO$_{2}$) showed very high sensitivity to conduction-inducing terminations (O, O, and F, respectively) with even minuscule amounts ($\approx$1\%) causing the materials to become conductive. This high sensitivity of the band gap to surface terminations may offer an explanation for the challenges in synthesizing semiconducting forms of MXenes.}

\end{tabular}

 \end{@twocolumnfalse} \vspace{0.6cm}

  ]

\renewcommand*\rmdefault{bch}\normalfont\upshape
\rmfamily
\section*{}
\vspace{-1cm}


\footnotetext{\textit{$^{*}$~Department of Physics, Faculty of Science, University of Ostrava, 30.~dubna 22, 701 03 Ostrava, Czech Republic, frantisek.karlicky@osu.cz}}

\footnotetext{\dag~Electronic Supplementary Information (ESI) available:
accuracy of lattice constants, bond dissociation energies, data showing chemical independence of terminal groups on opposite sides in MXenes with mixed termination, and the effect of terminal group patterns on the cohesion energy, band gaps, and band structures for semiconducting MXenes. See DOI: 00.0000/00000000.}



\section{Introduction}
\label{intro}
\begin{figure}[ht]
    \centering
    \includegraphics[width=5cm]{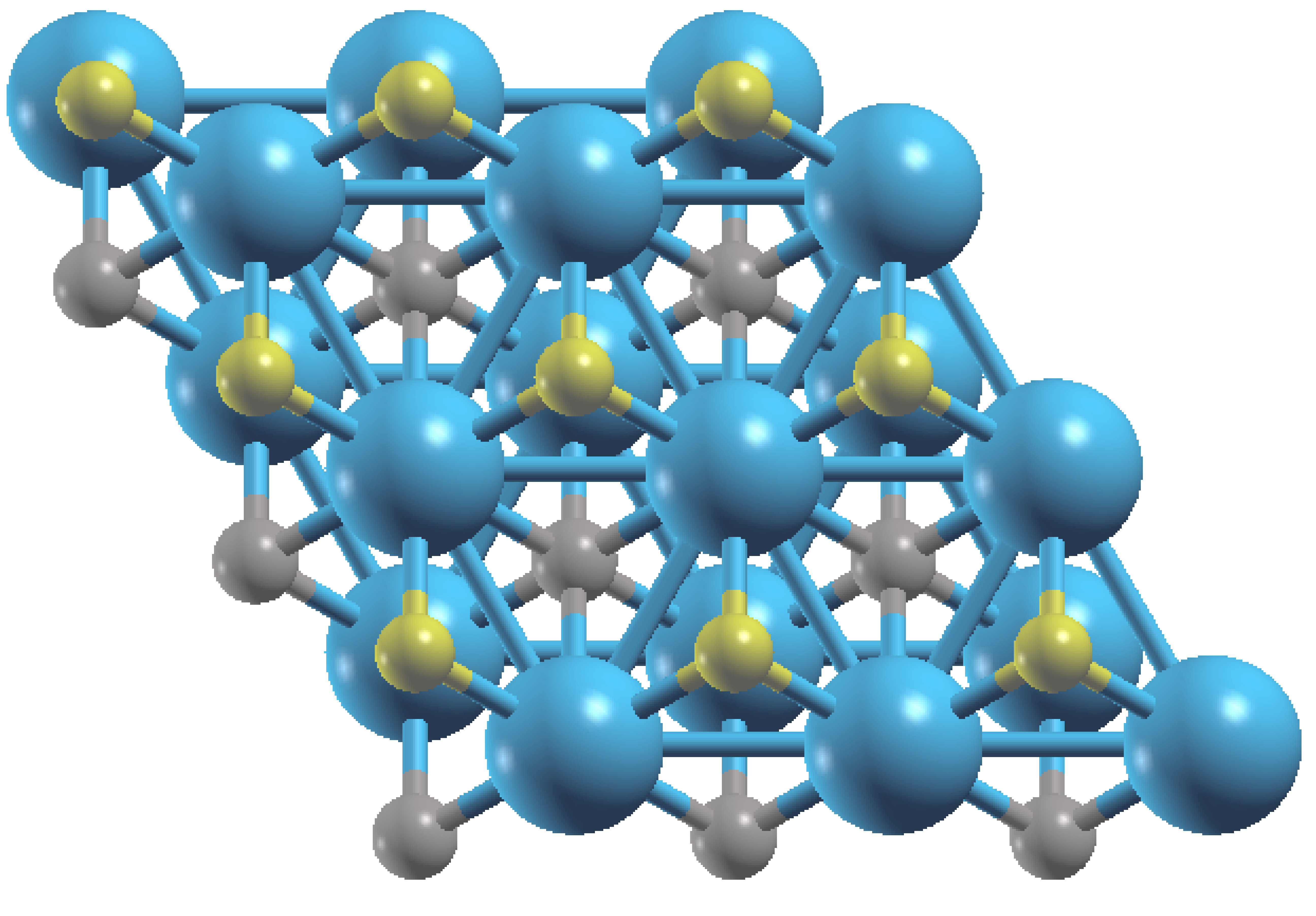}
    \includegraphics[width=3.3cm]{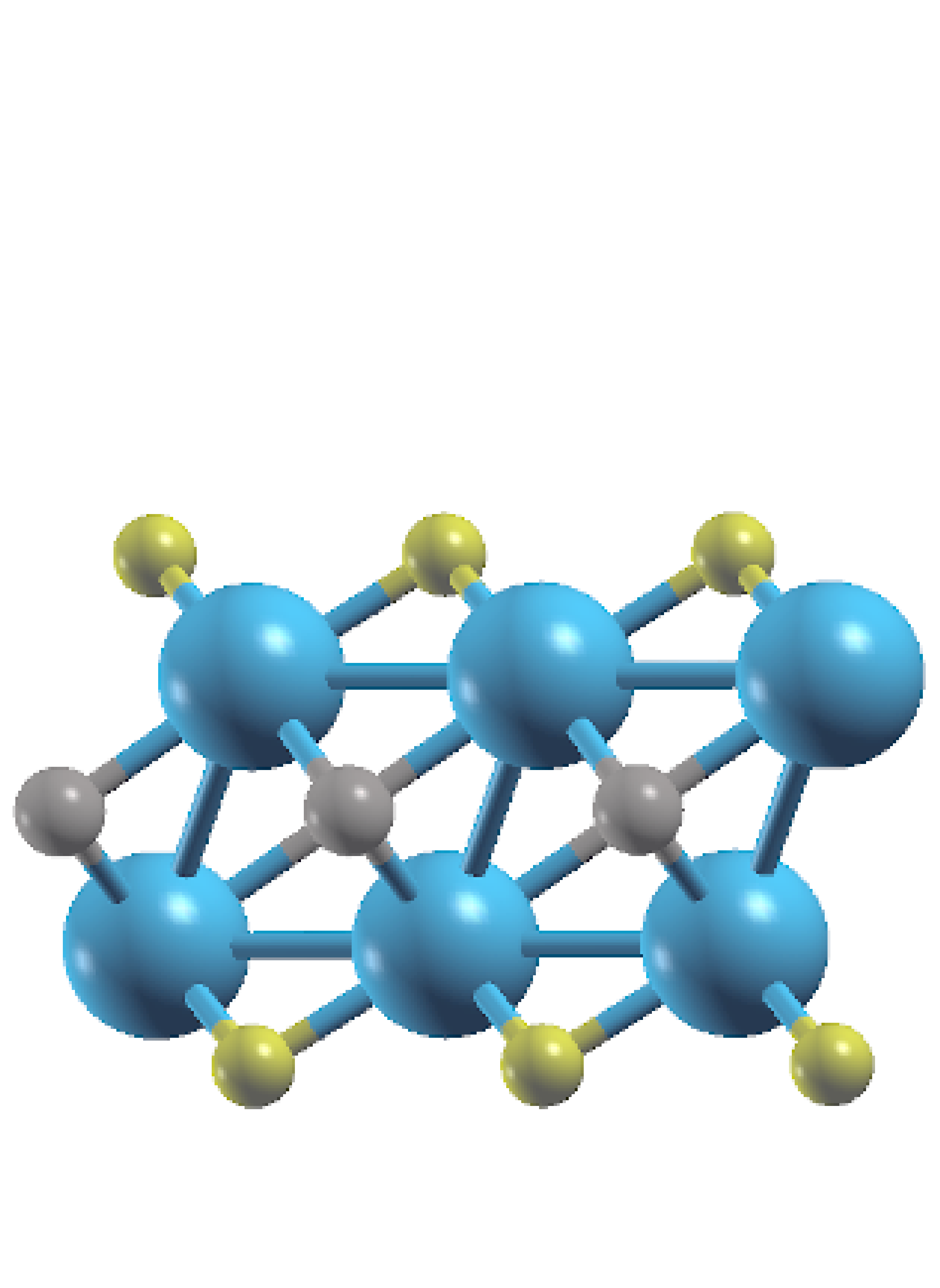}
    \caption{Top and side view of a 3x3 supercell of Sc$_{2}$CF$_{2}$ structure. Blue balls represent metal atoms Sc, grey carbon atoms, and yellow F-terminal groups.}
    \label{fig:struct}
\end{figure}

The recent rise of 2D material science has brought forth many promising nanomaterials. Some of the more recent ones are 2D transitional metal carbides, generally called MXenes. Structured in a hexagonal lattice with a composition of M$_{\mathrm{n+1}}$X$_{\mathrm{n}}$, where M is a transition metal (Sc, Ti, V, Cr, Hf, etc.) and X is either carbon or nitrogen.~\cite{Anasori2022, Shrabani2022} MXenes also possess a surface layer of terminal groups (T) on either side of the 2D plane, which generally depends on the preparation process and/or subsequent chemical modifications. While early computational studies have omitted these surface terminations in their models, it has been repeatedly proven that these T-groups play a pivotal role in determining the properties of the whole MXene.~\cite{Khanal2023,Brette2023,Wang2021} 
The surface of MXenes can be terminated by different species, ranging from halogen, oxygen, and hydroxyl terminations
through sulfur and selenium-based terminations~\cite{Tang2021}
to organic radicals.~\cite{Anasori2022}
An example structure of Sc$_{\mathrm{2}}$CF$_{\mathrm{2}}$ is shown in Fig.~\ref{fig:struct}.

Due to the many different combinations of M, X, and T as well as the possible mixing of metal atoms,
 MXenes exhibit a large compositional variety. 
In this study, we will be focusing on single-metal carbide MXenes. Even by limiting our models to this small island of the compositional space we can obtain materials with significantly different properties. In these 2D carbides, the terminal groups play a significant role in determining the band gap,~\cite{Ketolainen2022} work function,~\cite{Khazaei2017} hydrophilic behavior,~\cite{Hart2019} and catalytic activity.~\cite{Handoko2020} Experimental studies have reported different surface compositions for different metals. However quantitative characterization remains problematic due to the presence of surface contaminants from etching reagents and solvents, as well as the fact that T-groups are mostly light elements. Based on the preparation method, Ti-based MXenes have been reported to have only O and OH terminations~\cite{Urbankowski2016} or all three (O/OH/F).~\cite{Halim2016}  The question of quantitative surface composition has been addressed by several experimental~\cite{Hope2016} as well as theoretical studies,~\cite{Hu2018, Ibragimova2021} but due to the complexity of the surface composition, these works focus on one or few  M-elements with more or less homogeneous surfaces.

While previous studies have looked at MXenes where the surface terminations on opposite sides are not identical (Janus structures)\cite{Li2021,Jin2018,Frey2019}, a thorough examination on how they affect each other's chemistry is lacking. Our study addresses specifically this gap in our knowledge of MXenes by closely examining when the two opposite sides of functional groups have different compositions and/or patterns and how this affects the overall stability of the MXenes in question, their electronic properties, and the terminal group binding energies. We focus on the first three MXenes from the periodic table: Sc$_{2}$C, Ti$_{2}$C, and V$_{2}$C. While the equilibrium T-group composition of Ti$_{2}$C and V$_{2}$C has been discussed by other works, ~\cite{Ibragimova2021,Caffrey2018} an \textit{ab-initio} analysis of equilibrium surface functionalization of Sc$_{2}$C has not been conducted. It is worth mentioning that although the surface composition of Sc$_{2}$C has not been investigated, works already exist examining the chemical properties of Sc$_{2}$C with mixed surface composition~\cite{Yadav2022,Cui2024,Cui2023,Kumar2016,Sakhraoui2024}.

While the most prominent feature of MXenes is that they are hydrophilic 2D metals, in some specific combinations of d-elements and functional groups the band gap can open. While for most MXenes it is relatively well known when this happens, most research (as said previously) focused on models with a single type of T-group.
In our study, we provide the first look at the effect of gradual concentration changes of mixed terminations on the opening and size of Sc$_{2}$C band gap. This work also provides a rigorous analysis of the changes in the band-structure induced by terminal group substitutions in semiconducting forms of the studied MXenes as well as mapping of equilibrium T-group composition for Sc$_{2}$C, Ti$_{2}$C, and V$_{2}$C.

\section{Methods}
\label{methods}
All calculations have been performed using the periodic density functional theory (DFT) code Vienna Ab-initio Simulation Package (VASP)~\cite{Kresse_PhysRevB_47_1993, Kresse_PhysRevB_49_1994, Kresse_Comp_Mat_Sci_6_1996, Kresse_PhysRevB_54_1996} in version 6.4.2. The Perdew, Burke, and Ernzerhof (PBE) exchange-correlation functional in the generalized gradient approximation proposed by Perdew \textit{et al.}~\cite{Perdew_PhysRewLet_77_1996, Perdew_PhysRewLet_78_1997} was used as well as the meta-GGA strongly constrained and appropriately normed (SCAN) functional.~\cite{SCAN2} Lattice constants, structural relaxations, bond dissociation energies, cohesion energies, and band structures were calculated using the SCAN density functional. PBE was employed to generate a pre-converged wave function which was then used as a starting point in the SCAN-SCF calculation. The Kohn-Sham equations have been solved variationally in a plane-wave basis set using the projector-augmented-wave (PAW) method of Blöchl,~\cite{Blochl_PhysRevB_50_1994} as adapted by Kresse and Joubert.~\cite{Kresse_PhysRevB_59_1999} The convergence of the total energy of the cell with respect to the plane wave cutoff was set to 400 eV. The electronic self-consistency cycle convergence criterion was set to $\mathrm{10^{-8}}$ eV/cell. Structural relaxations employed quasi-Newton RMM-DIIS algorithm~\cite{Pulay1980} with an ionic step relaxation threshold of 10$^{-5}$ eV/\AA{}.

The Brillouin zone sampling was done by a gamma-centered mesh of 24$\times$24$\times$1  k-points for the unit cells of MXenes. For calculations of bond dissociation energies (BDE) in pure terminated structures we used a 3$\times$3$\times$1 supercell and an 8$\times$8$\times$1 k-point grid to maintain constant k-point density. In both unit cell calculations and BDE calculations we employed the tetrahedron electronic smearing method. 
For the calculation of mixed T-group composition, we used a 3$\times$3 supercell, sampled by a 3$\times$3$\times$1 k-point grid with Gaussian electronic smearing in relaxation runs and tetrahedron smearing in energy calculations. The accuracy of the k-point down-sampling has been verified and is shown in Fig.~S1. 

The z-direction cell sizes were set to 30~\AA{} making a vacuum layer of $\approx$ 25~\AA{}. While several different possible conformations of the MXene structure have been reported, for this study we focused on only the one shown in Fig.~\ref{fig:struct} (trigonal M$_2$C structure with T in hollow position) as it is reported to be the energetically most favorable one for almost all types of MXenes.~\cite{Dubeck2023,Zhang2021,Kumar2024, Khazaei2012}

As a measure of stability and energetic preference of a terminal group composition, we have chosen to evaluate the cohesion energies of the studied structures. The cohesion energy given by eq.~\ref{eq:coh} was calculated as the difference between the total energy of an MXene ($E_{MX}$) and the sum of atomic energies of its constituent elements ($E_{i}$) per atom, where $\nu_{\mathrm{i}}$ is the stochiometric coefficient of i-th element,
\begin{equation}
    E_{coh}= \frac{E_{MX}- \sum_{i} \nu_{i} E_{i}}{\sum_{i} \nu_{i}}.
    \label{eq:coh}
\end{equation}
For quantifying the effect of changing patterns and concentrations on individual terminal groups, we evaluate their bond dissociation energies as given by eq.~\ref{eq:bde},

\begin{equation}
    BDE=E_{MX}-E_{MX_{T-vacancy}}-E_{T},
\label{eq:bde}
\end{equation}
where $E_{MX_{T-vacancy}}$ is the energy of a system with a T-group vacancy and E$_{\mathrm{T}}$ is the energy of a free-standing terminal group calculated as a free-standing O or F atom or un-relaxed OH radical. Calculations of E$_{\mathrm{MX}}$ for all non-defective systems were performed without spin polarization since previous studies showed that the ground states of these MXenes are non-magnetic.~\cite{Sakhraoui2022, Ketolainen2022} Calculation of isolated terminal groups and MXenes with T-group vacancies were performed with spin polarization taken into account. To evaluate changes in the charge density we have employed the Bader charge analysis using the Bader code.~\cite{bader} For obtaining unfolded band structures of supercells we employed the  VaspBandUnfolding 
toolset~\cite{BandUnfol} which implements the methodology outlined by Popescu et al.~\cite{Popescu2012}

\section{Results and Discussion}
\label{results}

\subsection{MXenes with a single type of surface termination}

Lattice constants of MXenes with a single type of termination were fully optimized and are compiled in Tab.~\ref{tab:lat}. The reported lattice constants show a clear trend of shortening from Sc to V which is in accordance with the well-known fact of decreasing atomic radii of d-block elements. While the terminal groups affect the lattice constants to a lesser degree, accurate calculations of these values are essential since even small strains in the structure can significantly affect its band gap.~\cite{Ketolainen2022}

\begin{table}[ht]
\centering
   \caption{Lattice constant \textit{a}, interatomic distance ($d$) and inter-layer ($d_{z}$) distances between the various atomic species of M$_{2}$CT$_{2}$ where M= Sc, Ti, V. All reported values are in \AA{}. In the case of OH terminations the two numbers indicate O and H distances respectively.}
    \label{tab:lat}
    \small
\begin{tabular}{llllll}
\hline             
MXene      & $a$    & $d$(C-M) & $d_{z}$(C-M) & $d_{z}$(C-T) & $d_{z}$(M-T) \\
\hline 
\hline 
Sc$_{2}$CF$_{2}$    & 3.18 & 2.24   & 1.28         & 2.44         & 1.16         \\
Sc$_{2}$CO$_{2}$    & 3.16 & 2.55   & 1.78         & 2.55         & 0.76         \\
Sc$_{2}$C(OH)$_{2}$ & 3.23 & 2.26   & 1.27         & 2.53/3.49    & 1.25/2.22    \\
\hline 
Ti$_{2}$CF$_{2}$    & 2.95 & 2.06   & 1.16         & 2.42         & 1.26         \\
Ti$_{2}$CO$_{2}$    & 2.98 & 2.16   & 1.31         & 2.22         & 0.91         \\
Ti$_{2}$C(OH)$_{2}$ & 3.00 & 2.08   & 1.16         & 2.44/3.40    & 1.28/2.24    \\
\hline 
V$_{2}$CF$_{2}$     & 2.87 & 1.96   & 1.05         & 2.34         & 1.29         \\
V$_{2}$CO$_{2}$     & 2.85 & 2.03   & 1.19         & 2.21         & 1.02         \\
V$_{2}$C(OH)$_{2}$  & 2.92 & 1.98   & 1.04         & 2.32/3.28    & 1.27/2.24    \\
\hline
\end{tabular}
\end{table}

In the purely terminated structures (Tab.~\ref{tab:lat}), we observed that the distance between the carbon and terminal group layers ($d_{z}$(C-T)) vary only slightly ($\approx$0.1~\AA{}). This holds also true in the case of $d_{z}$(C-O) distances in both O and OH. This is somewhat surprising since, as we will show later, O and OH terminations exhibit significantly different bonding behavior. Minimal variance is also found in the carbon to metal layer distances ($d_{z}$(C-M)) for F and OH terminations. On the other hand, the O groups tend to pull the metal atom layer further away from the carbon layer compared to F and OH terminations. The fact that the whole metal layer is pulled away, rather than the change occurring in the metal atoms lattice, is supported by the elongation of $d$(C-M) distances.  In Ti$_{2}$C and V$_{2}$C this effect is relatively small (0.1~\AA{} and 0.2~\AA{}) but in the case of Sc$_{2}$C the Sc atoms are pulled up to 0.5~\AA{} away from the carbon layer.

As a first look into the effect of T-groups on opposite sides, we performed a Bader charge analysis and calculated the BDE of individual terminal groups in a set of model systems, in which the effects of opposite sides of the MXene are most prominent. These model systems have both sides fully covered by one type of terminal group. The results of the Bader charge analysis calculated for all three metals in their respective primitive cells are summarized in Tab.~\ref{tab:bader}. The table shows the effective partial charges/oxidation states on all atomic species in the MXene. 

\begin{table}[ht]
\centering
   \caption{Effective partial charges/oxidation states (in $e$) of atomic species in M$_{2}$CTT' where M/T denotes terminal groups on the top of the MXene and M'/T' on the bottom.}
    \label{tab:bader}
\begin{tabular}{lrrrrlr}
\hline
 & \multicolumn{1}{c}{FF}     & \multicolumn{1}{c}{FO}    & \multicolumn{1}{c}{F(OH)}   & \multicolumn{1}{c}{OO}     & \multicolumn{1}{c}{(OH)(OH)}    & \multicolumn{1}{c}{O(OH)}   \\
\hline
\hline 
C     & -2.09 & -1.70 & -2.09 & -1.31 & -2.10 & -1.69 \\
Sc    & 1.85  & 1.84  & 1.85  & 1.94  & 1.82  & 1.94  \\
Sc’   & 1.85  & 1.94  & 1.83  & 1.94  & 1.83  & 1.82  \\
T     & -0.81 & -0.81 & -0.82 & -1.29 & -1.37 & -1.28 \\
T’    & -0.81 & -1.27 & -1.38 & -1.29 & -1.37 & -1.36 \\
H     & -     & -     & 0.59  & -     & 0.59  & 0.58  \\
\hline
C     & -1.97 & -1.86 & -1.96 & -1.75 & -1.98 & -1.86 \\
Ti    & 1.75  & 1.78  & 1.76  & 2.03  & 1.78  & 2.02  \\
Ti’   & 1.75  & 2.01  & 1.75  & 2.03  & 1.78  & 1.78  \\
T     & -0.76 & -0.77 & -0.77 & -1.16 & -1.36 & -1.17 \\
T’    & -0.77 & -1.16 & -1.36 & -1.16 & -1.34 & -1.34 \\
H     & -     & -     & 0.59  & -     & 0.57  & 0.57  \\
\hline
C     & -1.75 & -1.68 & -1.74 & -1.62 & -1.75 & -1.68 \\
V     & 1.61  & 1.63  & 1.61  & 1.84  & 1.61  & 1.85  \\
V’    & 1.63  & 1.83  & 1.61  & 1.84  & 1.62  & 1.62  \\
T     & -0.74 & -0.76 & -0.74 & -1.03 & -1.32 & -1.04 \\
T’    & -0.74 & -1.03 & -1.34 & -1.03 & -1.33 & -1.32 \\
H     & -     & -     & 0.59  & -     & 0.58  & 0.58  \\
\hline 
\end{tabular}
\end{table}

As expected, based on differences in electronegativity, in Tab.~\ref{tab:bader} the metal atoms show a positive oxidation state of $\approx$+2, F-terminations $\approx$-1, oxygen atoms $\approx$-1 in both O and OH groups and hydrogen $\approx$-0.5. The observed oxidation states of the metal atoms are in accordance with previous works.~\cite{Yusupov2023} and while not completely comparable, similar trends were observed for   Janus type Ti$_{3}$C$_{2}$FO by Li et al.~\cite{Li2021}. While commonly the oxidation state of carbon atoms in ionic carbides (methides) is ascribed to be -4,~\cite{Atkins2009-pq} we show that in the case of our studied MXenes, it is $\approx$-2. A somewhat surprising observation from Tab.~\ref{tab:bader} is that the terminal groups are draining electrons from the metal and carbon atoms at the same time rather than just the metal atom. This is most prominent when one side is fully covered by F-terminations and the other by O. One would expect the carbon to compensate for this drain from the less electronegative metal atom on the other side but this does not happen. One can therefore think of the carbon layer as a form of electron sink for the terminal groups. The strongest electronic drain is exhibited by the O groups with F and OH showing almost identical C and M oxidation states. This is due to the oxygen atom in OH terminations compensating its electron drain from the hydrogen. These observations are true for all three metals with the the effect of electronic drain decreasing with the increasing electronegativity in the order Sc>Ti>V. 
The charge analysis indicates that the terminal groups on opposite sides of the MXene, do not affect each other, and neither do they affect the opposite metal layer to a relevant degree.

Another parameter that was analyzed, that could shed light on the interplay between terminal groups on opposite sides is the bond dissociation energy, which is listed in Tab.~\ref{tab:BDE}. The reported BDEs were calculated using a 3$\times$3 supercell with one terminal group removed. For MXenes with F and OH-group vacancies, one nonmagnetic (magnetic moment of 0~$\mu_{B}$) and one ferromagnetic (1~$\mu_{B}$) state was calculated. In the case of oxygen vacancies, one nonmagnetic state and two ferromagnetic states (1~$\mu_{B}$ and 2~$\mu_{B}$) were calculated. Tab.~\ref{tab:BDE} shows the BDE calculated from the energies of the defected systems (E$_{MX_{T-vacancy}}$) with the lowest-lying magnetic states. BDEs for all systems and magnetic states are shown in Tab.~S2.

\begin{table}[ht]
\centering
   \caption{Bond dissociation energies for F, O, and OH terminal groups calculated in 3x3 M$_{2}$CTT' supercells, where T denotes terminal groups on the top of the MXene and T' on the bottom.}
    \label{tab:BDE}
\begin{tabular}{lcccc}
\hline 
Coverage    &  \multirow{2}{4em}{Dissociated bond}        & \multicolumn{3}{c}{BDE {[}eV{]}} \\
TT' &  & Sc        & Ti        & V        \\
\hline
\hline 
FF       & -F  & -7.78 & -6.57 & -5.64 \\
(OH)F    & -F  & -7.72 & -6.57 & -5.64 \\
OF       & -F  & -6.55 & -6.58 & -5.20 \\
FO       & -O  & -7.18 & -7.76 & -6.83 \\
(OH)O    & -O  & -7.40 & -7.76 & -6.81 \\
OO       & -O  & -7.52 & -8.60 & -6.56 \\
F(OH)    & -OH & -6.89 & -6.02 & -5.20 \\
(OH)(OH) & -OH & -6.77 & -6.05 & -5.16 \\
O(OH)    & -OH & -5.99 & -5.95 & -4.71 \\
\hline
\end{tabular}
\end{table}

The bond dissociation energy of a terminal group is directly proportional to its bond strength and the group's contribution to the cohesion energy and overall stability of the material. The calculated BDEs are relatively high (4.7-8.6 eV) and comparable with very strongly bonded covalent systems (BDE$_{O_{2}}$=5.16~eV, BDE$_{C_{2}^{sp2}}$=7.4~eV). A possible explanation and better model of the surface group chemistry is to look at the interaction as three metal-T bonds rather than just a single surface-to-T-group bond. A thorough investigation of the nature of these bonds could shed even more light on the complex chemistry of T-groups.

In general, the BDEs of specific terminal groups vary only slightly with different counterparts on the other side. An exception to this is structures that have a different conducting/semiconducting nature in a given series. For Sc-MXenes, the BDE of F in Sc$_{2}$COF and BDE of OH in  Sc$_{2}$CO(OH) are $\approx$1~eV lower than than the BDEs in other compositions. Sc$_{2}$CO(OH) and Sc$_{2}$COF are conductive while the other structures in their respective series are semiconducting, showing that the terminal groups in semiconducting MXenes have stronger bonds with the bare MXene. This generalization is supported by the BDE of O in Ti$_{2}$CO$_{2}$ where it is $\approx$1~eV higher compared to other structures and it is the only semiconducting Ti-based MXene that we have studied. For Ti-MXenes and V-MXenes, we have found the order of bond strength to be $BDE_{O} > BDE_{F} > BDE_{OH}$.  For Ti-MXenes, this order has also been reported by Hu et al.~\cite{Hu2018}. For Sc-MXenes the general order of BDEs is $BDE_{F} > BDE_{O} > BDE_{OH}$, but if we take into account only the conducting forms (containing oxygen) the order switches to $BDE_{O} > BDE_{F} > BDE_{OH}$. 

Even when changing the whole opposite side of the MXene to a different termination the BDE and the atomic charge densities/oxidation states remained almost unaffected. This data leads to the conclusion that both sides of the MXene behave like chemically distinct environments and that the terminal groups do not affect each other through the carbide layer.

\subsection{MXenes with mixed surface termination}

To reduce lattice strain in the mixed termination structures we approximate the lattice constant as a weighted average from pure terminated structures with weights representing the number of terminal groups on both sides of the MXene. To check the validity of this approximation, we performed full structural relaxation of three random mixed terminations for Sc-MXenes and compared the results with the weighted average structure see Tab.~S1 The results show that a weighted average of lattice constants reasonably predicts equilibrium lattice parameters with only a negligible strain on the cell. Thus mixed structures with weighted average lattice constants should be an adequate model for assessing the T-group composition of the studied MXenes. The atomic positions in all studied structures were fully optimized. The structural relaxations indicate that the atoms undergo relatively small displacements if the surface is composed of a mix of terminal groups. While these displacements were relatively small (less than 0.1~\AA{}), these displacements had a significant impact on the band gaps and BDEs when compared to unrelaxed structures. We have therefore carefully optimized the atomic positions in all studied mixed surface termination structures.

\begin{table}[ht]
\centering
  \caption{Carbon to metal layer distances. The presented numbers are average values calculated from the whole set of mixed termination structures. All reported values are in~\AA{}}
\label{tab:mixed_dist}
\begin{tabular}{lccc}
\hline
\multirow{2}{8em}{Nearest T-groups to Metal atom} & \multirow{2}{4em}{$d_{z}$(C-Sc)}           & \multirow{2}{4em}{$d_{z}$(C-Ti)}           & \multirow{2}{4em}{$d_{z}$(C-V)}           \\
 &&& \\
    \hline
    \hline 
FFF           & 1.28         & 1.16         & 1.05        \\
FFO           & 1.44         & 1.27         & 1.10        \\
FOO           & 1.56         & 1.19         & 1.15        \\
OOO           & 1.78         & 1.31         & 1.19        \\
FF(OH)        & 1.27         & 1.14         & 1.02        \\
F(OH)(OH)     & 1.28         & 1.15         & 1.05        \\
(OH)(OH)(OH)  & 1.27         & 1.16         & 1.04        \\
OO(OH)        & 1.55         & 0.94         & 0.85        \\
O(OH)(OH)     & 1.43         & 1.19         & 1.10        \\
\hline 
\end{tabular}
\end{table}

Structural analysis of mixed terminations showed significant differences in $d_{z}$(C-M) distances and negligible ones in $d_{z}$(C-T) depending on the pattern of terminal groups. We do not show d$_{z}$(M-T) changes since they are complementary to d$_{z}$(C-M) shifts due to $d_{z}$(C-T) remaining constant. The carbon layer remained unperturbed in all of the studied structures. The d$_{z}$(C-M) distances show a dependence only on the immediate neighboring terminal groups of the metal atom. The average distances based on the T-groups surrounding the metal atoms are summarized in Tab.~\ref{tab:mixed_dist}. The O-terminations in Sc-MXene pull the metal atom away from the carbon layer up to 0.5~\AA{} based on the increasing number of O-terminations in sites around the Sc atom. These shifts have been observed in Ti- and V-MXenes as well, whilst to a lesser degree. The maximum shift in Ti and V was 0.14~\AA{}. Somewhat counterintuitively the $d_{z}$(C-T) distances remained almost identical to those in pure terminated structures (Tab.~\ref{tab:lat}) regardless of how the metal atoms underneath were shifted. 

\begin{figure}[!ht]
	\centering
	\includegraphics[width=0.3\textwidth]{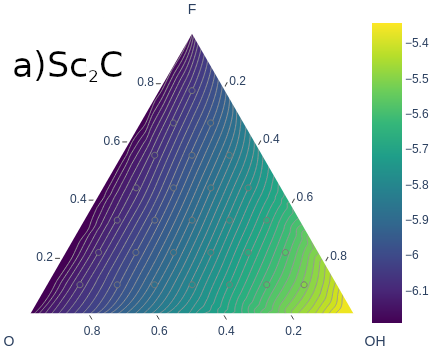}
    \includegraphics[width=0.3\textwidth]{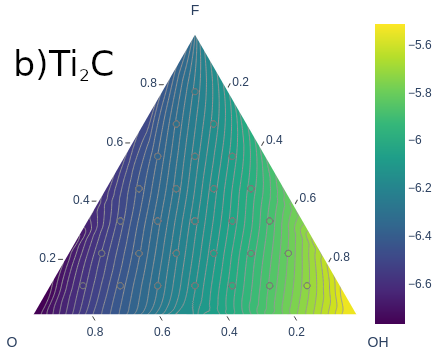}
    \includegraphics[width=0.3\textwidth]{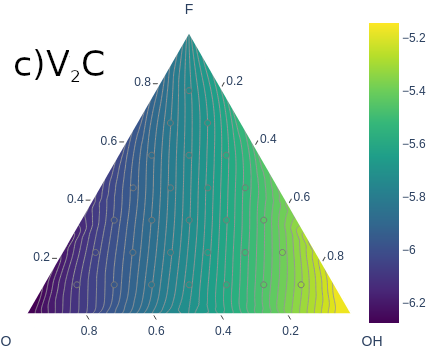}
	\caption{Cohesion energies [eV/atom] of M$_{2}$C (M=Sc, Ti, V) MXenes as a function of O, OH, and F terminal group coverage. The cohesion energies were calculated in a $3 \times 3$ super-cell with the pattern mirrored on both sides. Coverage is represented as a fraction of the number of terminal groups to all available sites on a single side. Lower cohesion energies indicate a more stable system and, therefore, a more probable composition. Each point represents one or more chemically distinct patterns with an identical T-group concentration but we only show the lowest recorded cohesion energy for a given surface composition.}
	\label{figure:Cohesion}
\end{figure}

\begin{figure}[!ht]
	\centering
    \includegraphics[width=0.40\textwidth]{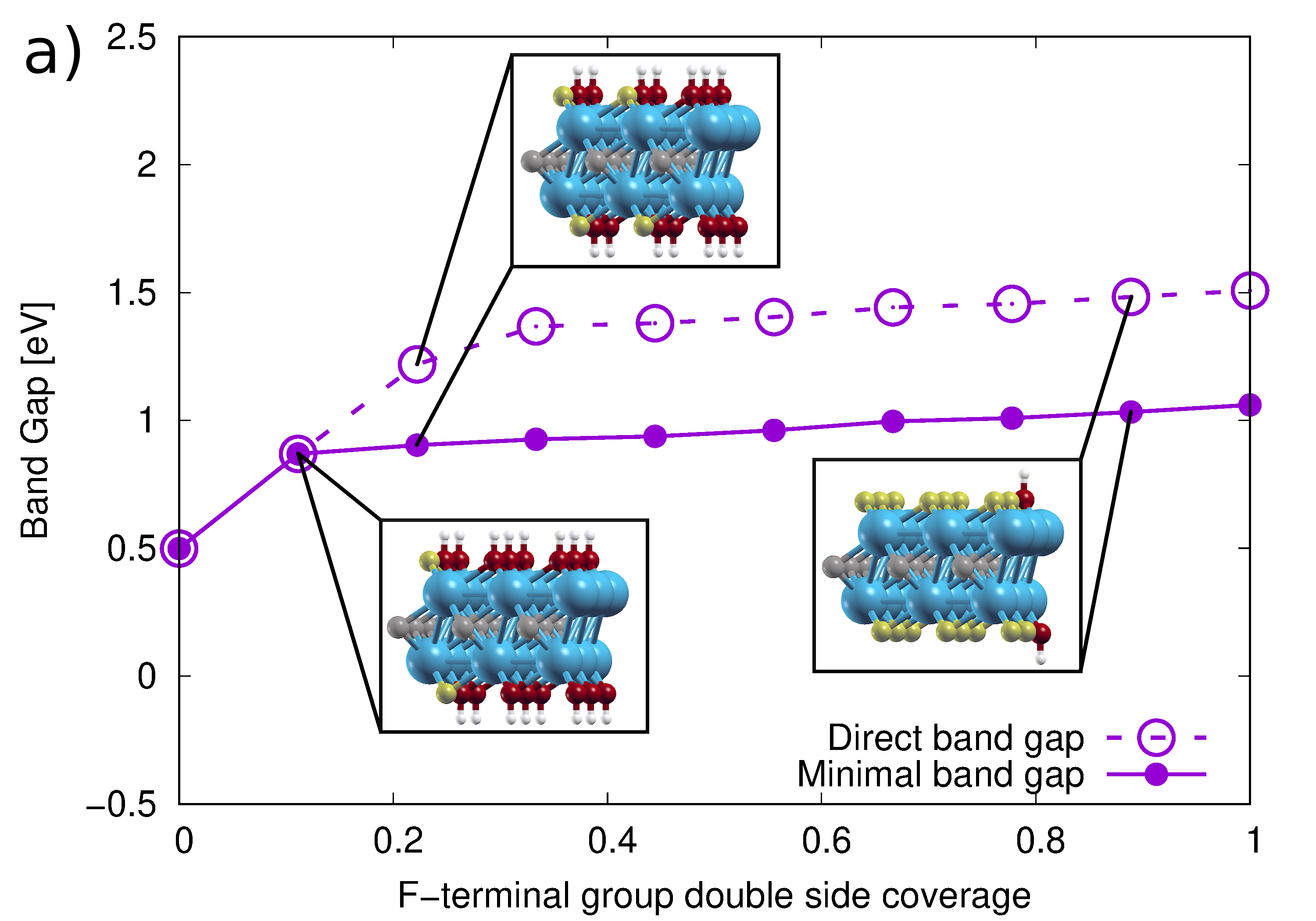}
    \includegraphics[width=0.40\textwidth]{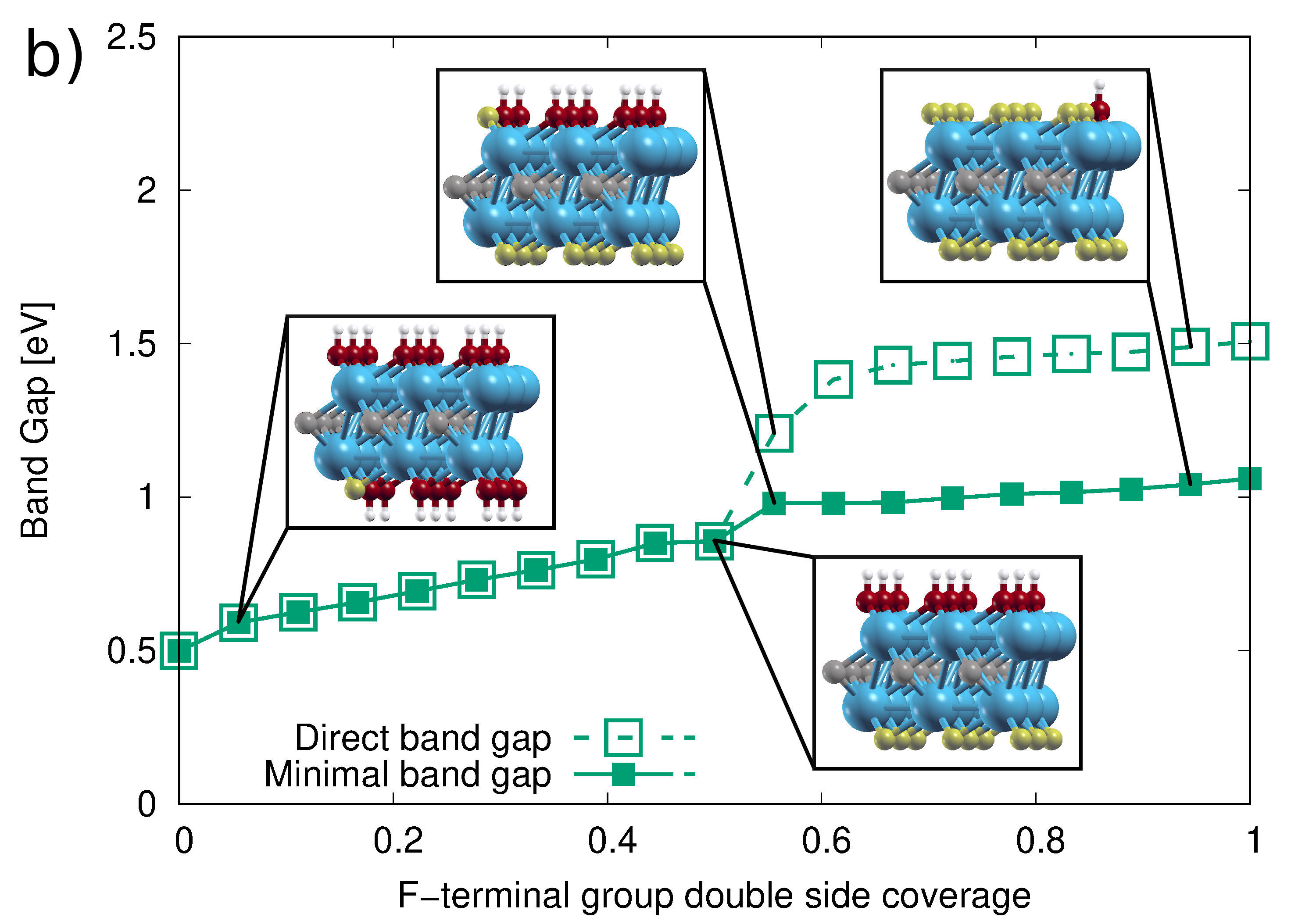}
	\caption{Band gap of Sc$_2$CF$_{2c}$(OH)$_{(2-2c)}$ as a function of F-group coverage ($0\le c \le 1$) represented as a fraction of the number of F-terminal groups to all available (both sides) sites. The graph represents a substation of OH-groups with F-groups: a) mirrored on both sides b) first on one side of the MXene and after $c$ = 0.5 continuing on the other side.}
	\label{fig:Sc_gap_con}
\end{figure}
\begin{figure*}[ht]
    \centering
    \includegraphics[width=0.24\textwidth]{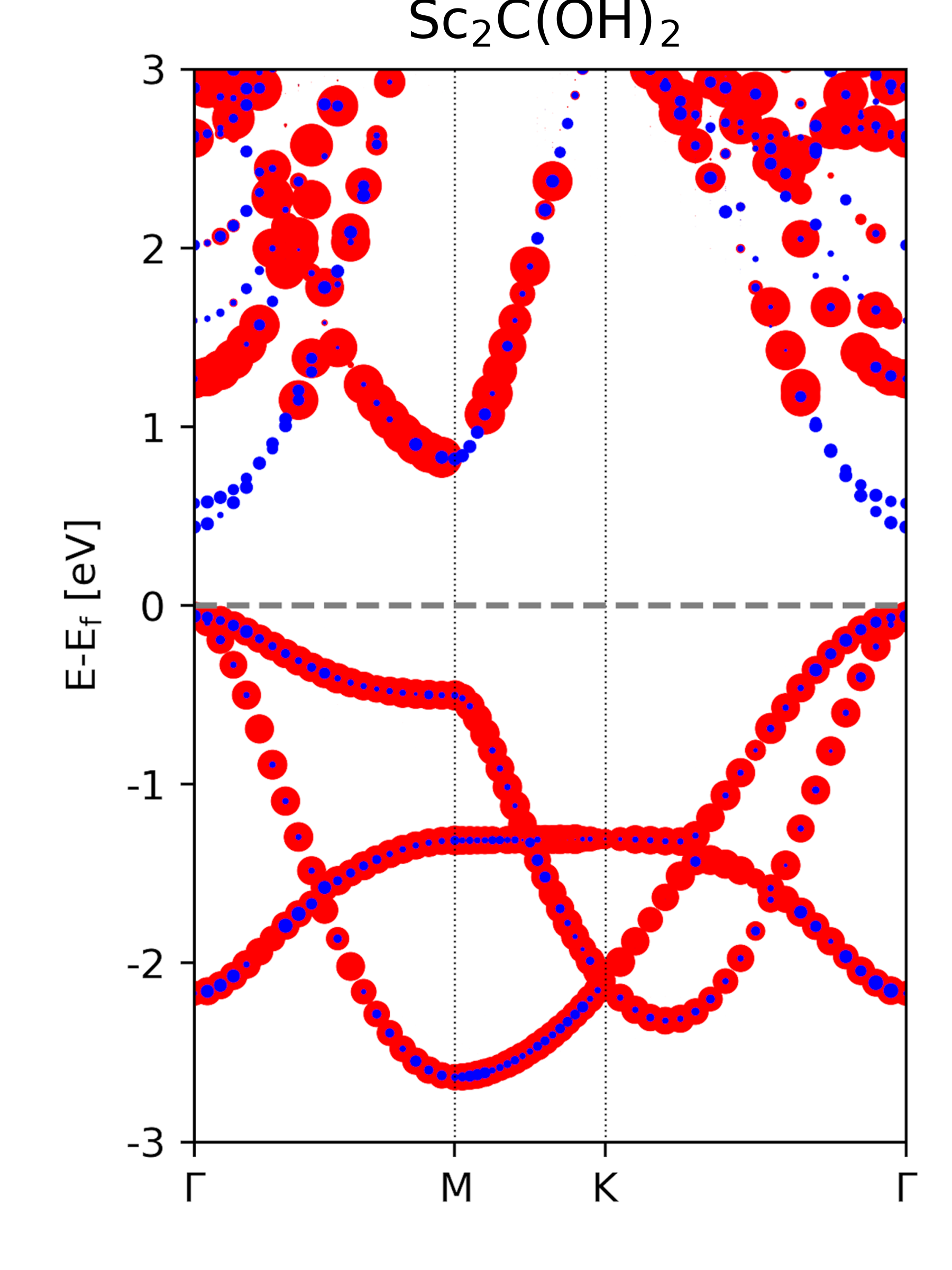}
    \includegraphics[width=0.24\textwidth]{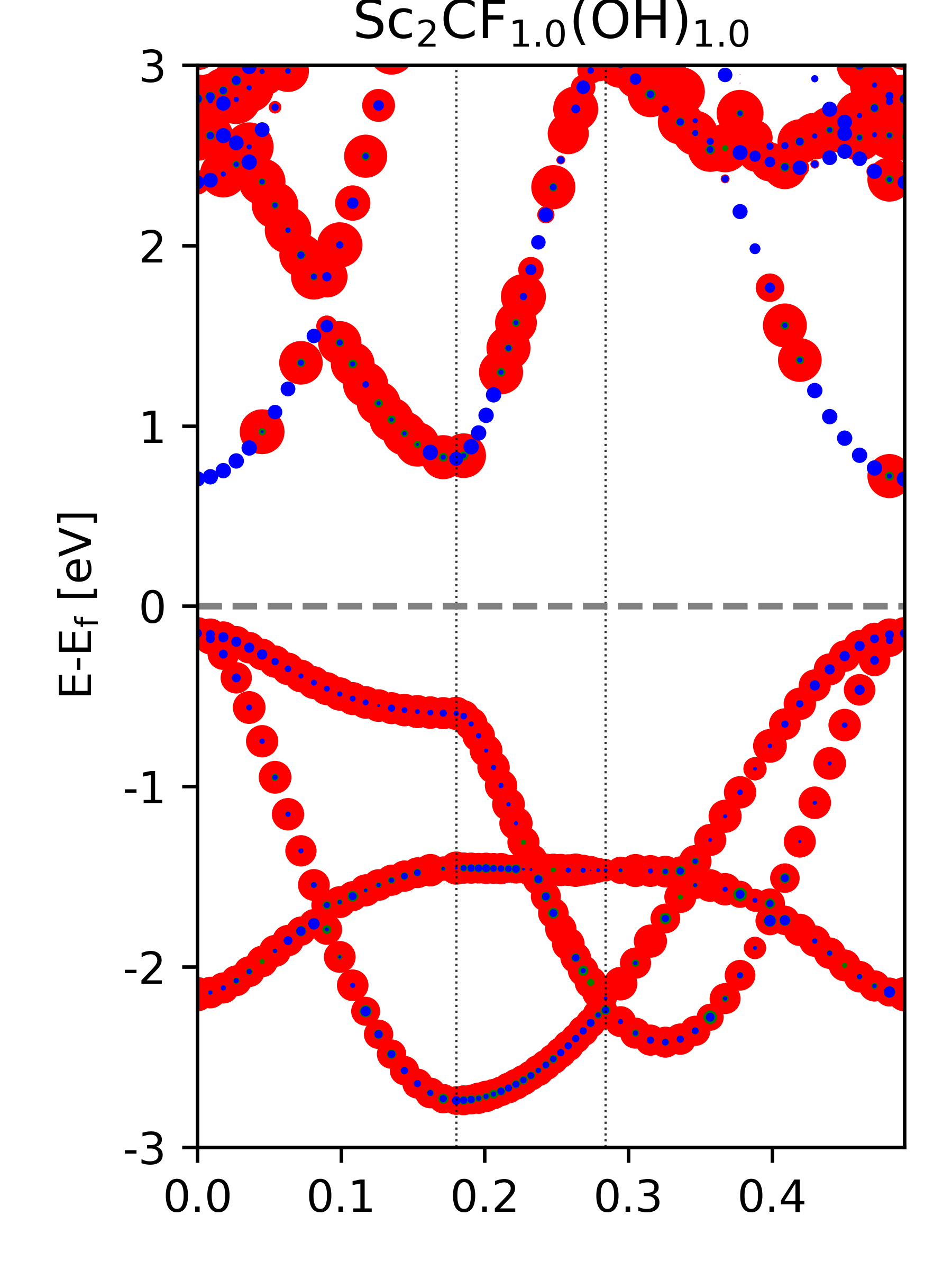}
    \includegraphics[width=0.24\textwidth]{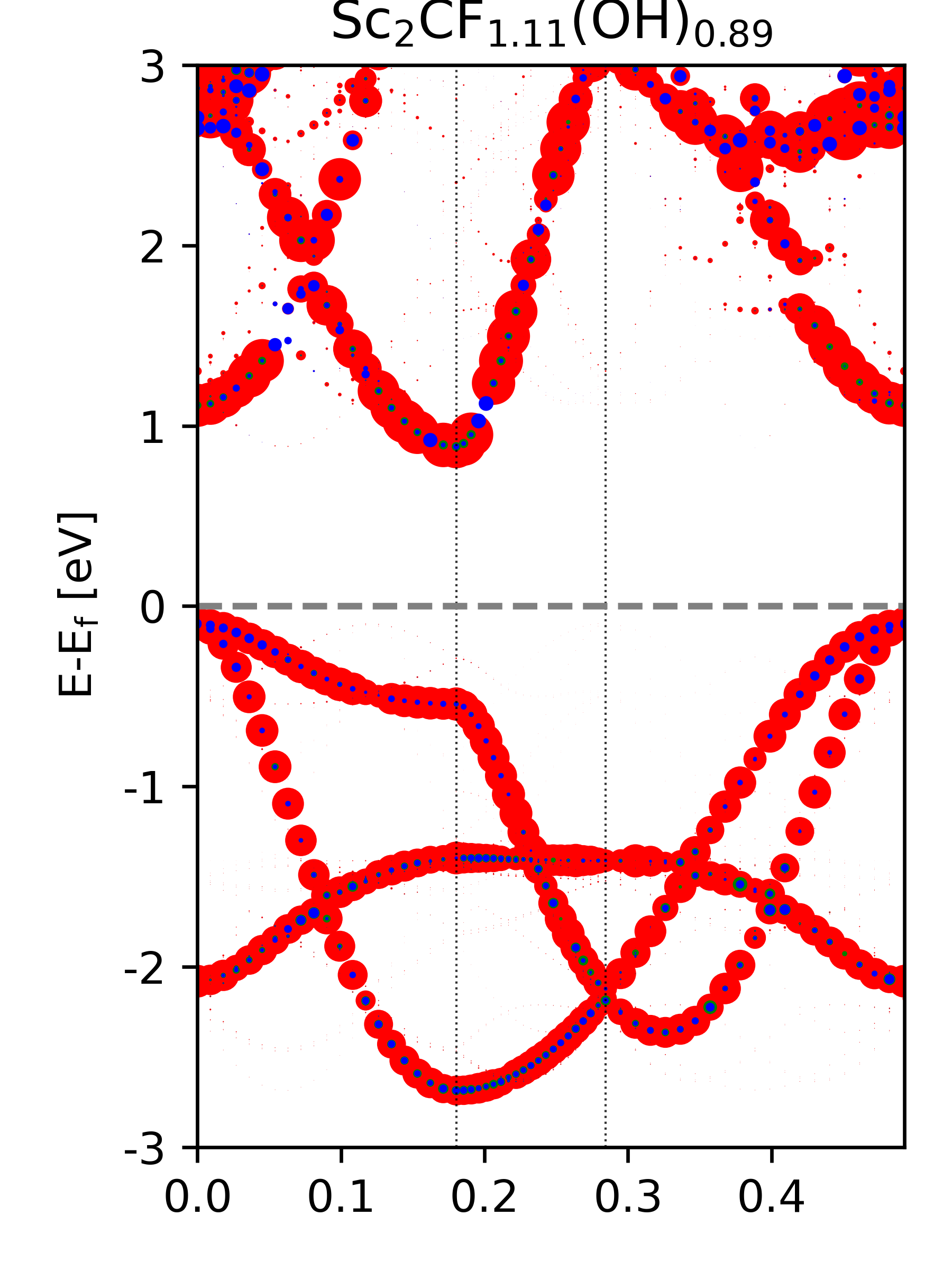}
    \includegraphics[width=0.24\textwidth]{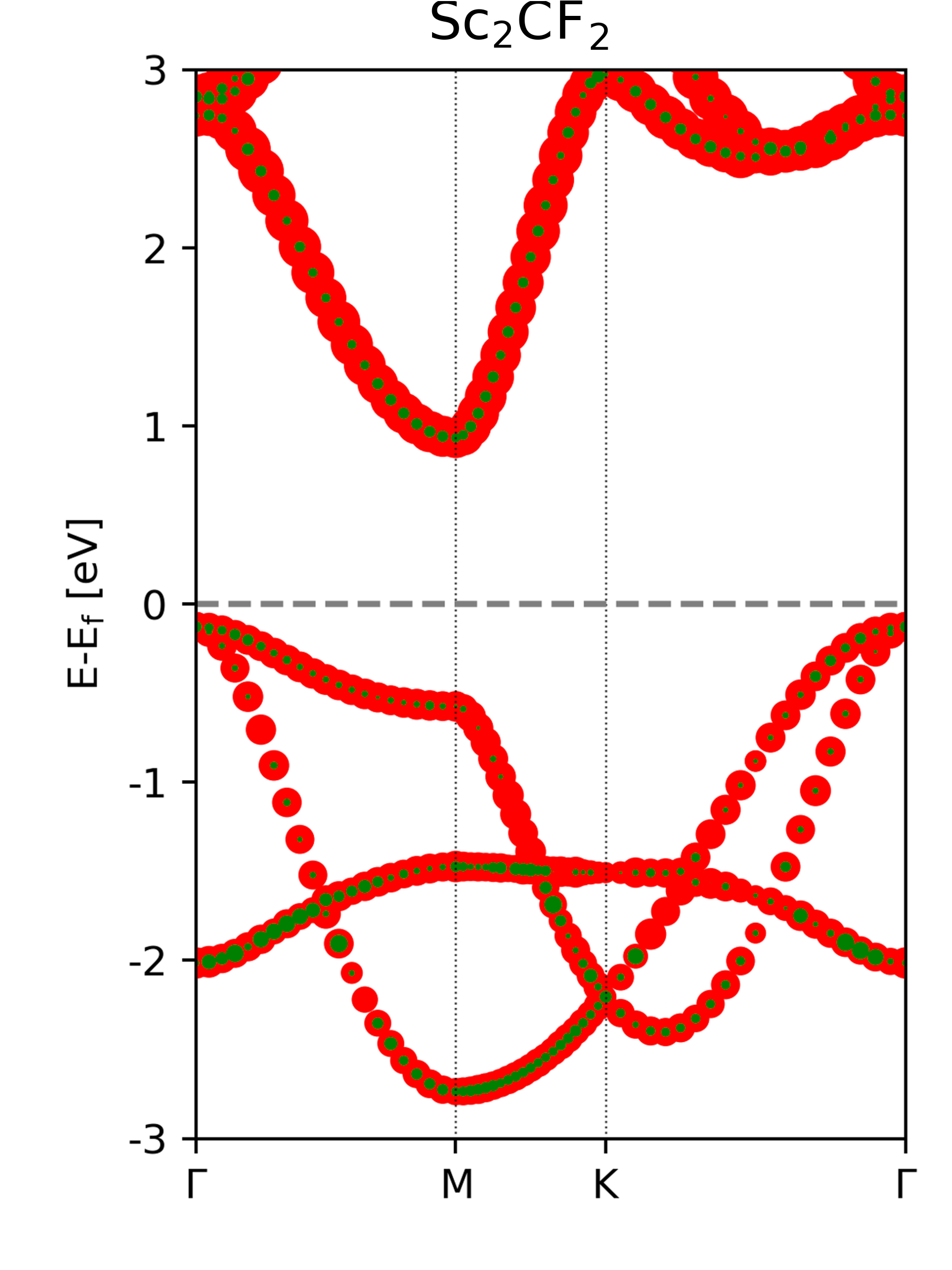}
    \caption{Atomically resolved band structures of non-mirrored Sc$_2$CF$_{2c}$(OH)$_{(2-2c)}$ with increasing F-group coverage ($0\le c \le 1$). The Fermi level was set to 0. Red circles correspond to band contributions from Sc atoms, green from F atoms, and blue from O atoms. The size of the dots corresponds to the spectral contributions of the atomic species to the band as outlined by Popescu et al.~\cite{Popescu2012}}
    \label{fig:FOH_bs}
\end{figure*}

While the order of BDEs in Tab.~\ref{tab:BDE} indicates which functional groups will be energetically preferred, it says nothing about the interplay between different T-groups when next to each other or arranged in specific patterns.  With a 3$\times$3 supercell, there are 9 terminal group positions for a single MXene side. When considering that these sites would be occupied by either F, O, or OH, this gives rise to more than forty thousand possible combinations of which only 627 are chemically unique patterns, meaning that they are invariant towards lateral translations and rotations within the hexagonal lattice. Previous calculations on pure surface terminations indicated that the opposite sides should be independent of each other. To have a direct proof of this, we have tested several mixed terminations (results shown in Tab.~S3 and associated Fig.~S2) and have confirmed that when keeping the composition and pattern constant on the top side while varying the pattern on the bottom,  the cohesion energy (and by extension BDE) and the band gap is independent of shifts and/or rotations of the bottom pattern. The variation on both $E_{coh}$ and band gaps was around 0.01 eV/atom and 0.01 eV respectively.

For all three metals (Sc, Ti, V) and all chemically unique patterns of terminal groups, we calculated the cohesion energies using the SCAN density functional. The pattern on the other side of the calculated structures was mirrored since as we have shown previously, terminal groups do not affect each other across the carbon layer. The results for the cohesion energies are summarized in Fig.~\ref{figure:Cohesion}. The cohesive energies for all structures are negative, indicating that they are thermodynamically stable.

A major find deduced from Fig.~\ref{figure:Cohesion} is that the pattern of terminal groups does not noticeably affect the cohesion energy. Each point in the ternary graph represents patterns with a specific concentration of the three terminal groups at which several chemically unique patterns can occur. In the graphs, we report only the lowest recorded cohesion energies for given concentrations. It was found that the cohesion energy of patterns with identical F, O, and OH concentrations vary only slightly for all three metals, on average in the range of 0.001-0.05 eV/atom. A more detailed depiction of pattern energy variation is shown in Fig.~S3 and ~S4. This indicates that there is no energetic reason for the formation of specific patterns, i.e., phase separation or agglomeration of the functional groups which agrees with previous findings~\cite{Ibragimova2019}. The predominant factor determining the cohesion energy is the T-group concentrations.  The cohesion energies of purely terminated structures are included in Tab.~S4 

The results for Ti$_{2}$C can be contrasted with the findings of Ibragimova et al.~\cite{Ibragimova2021} A common feature in both studies is the aversion of Ti$_{2}$C toward high concentrations of OH groups on the surface. While this paper predicts a clear preference for high surface concentrations of O-terminal groups Ibragimova et al.~\cite{Ibragimova2021} report a preference of Ti$_{2}$C to form surfaces with close to equal concentrations of O and OH. This difference is due to Ibragimova et al.~\cite{Ibragimova2021} accounting for a chemical environment with a pH=0 and calculating the Gibbs free energy as an indicator of composition, while our work is concerned only with system energetics in vacuum. This may also explain why, even though our work and Hu et al. report OH as the least favorable terminal group, other works such as ~\cite{Ibragimova2021} and experimentally prepared Ti-based MXenes exhibit non-negligible surface coverage of OH-groups.~\cite{Hope2016,Wang2015}


The order of most to least stable terminations for Ti-MXenes and V-MXenes is consistent with the order of  BDEs. Both metals have in common that the least favorable termination is the hydroxyl group and both show a clear preference towards pure oxygen-terminated surfaces. The jump in BDE of Ti$_{2}$CO$_{2}$ does not affect the cohesion energies since it doesn't change the bond strength order. Our findings for Ti-MXenes are in agreement with previous studies.~\cite{Hu2018,Ibragimova2021,Zhang2021} 
In the case of Sc$_{2}$C, the order of cohesion energies is the same as the order of BDEs for conductive structures. This was to be expected since Sc-based MXenes with mixed surface composition are almost exclusively conductive. Sc$_{2}$C in contrast to Ti$_{2}$C a V$_{2}$C seems to form an equilibrium composition of Sc$_{2}$CF$_{0.54}$O$_{1.46}$ (Fig.~\ref{figure:Cohesion} and detailed Fig.~S4) rather than a single T-group surface coverage.

\subsection{Band structures and band gaps.}

While the energetics of terminal groups show a preference for high concentrations of one terminal group, experimentally prepared MXenes exhibit a large variety in their surface composition~\cite{Fagerli2022, Hope2016} strongly dependent on the preparation method. The band gap of MXenes depends on the type of surface termination and while theoretical approaches have explored the effect of pure terminations on the band gap,~\cite{Ketolainen2022, Sakhraoui2022, Champagne2020}
no previous works have addressed the effect of mixed composition and the effect of different compositions on either side of the MXene. All discussed band gaps were calculated using the SCAN meta-GGA density functional.

When analyzing the effect of surface composition on the band gap we found that V$_{2}$C structures remained conductive regardless of any variation in their T-groups which is consistent with previous studies.~\cite{Khazaei2017,Zha2015}
Ti$_{2}$C structures exhibited a nonzero band gap only for the double-sided fully oxygenated structure,~\cite{Kumar2023} with a direct band gap of 0.758 eV (in $\Gamma$) and indirect 0.543 eV ($\Gamma \rightarrow M$). All other structures exhibited conductive behavior. We calculated the band structure of Ti$_{2}C$ with F and OH-group concentrations as low as Ti$_{2}$CO$_{1.97}$F$_{0.03}$ and Ti$_{2}$CO$_{1.97}$(OH)$_{0.03}$ (7$\times$7 supercell with one oxygen group substituted with F or OH) and the structures remained conductive. 
Similar observations were reported by Li et al.~\cite{Li2021} although for significantly higher concentrations of conduction inducing T-groups.

Sc$_{2}$C exhibits semiconducting behavior, opposite to Ti-based MXenes, with only F and OH terminations and the presence of even one O-termination on one side of the structure causes the material to become conductive. We tested this in a similar fashion to Ti$_{2}$C where we lowered the O-group concentrations down to Sc$_{2}$CF$_{1.97}$O$_{0.03}$ and Sc$_{2}$C(OH)$_{1.97}$O$_{0.03}$ (7$\times$7 supercell containing only one oxygen group) with the band gap still not opening. This transition is shown in Fig.~S5 for a 3$\times$3 supercell where the increasing number of oxygen terminations gradually shift the valence bands higher above the Fermi level slowly converging towards the band structure of Sc$_{2}$CO$_{2}$. This very high sensitivity of the gap closing with the mere presence of certain functional groups may explain the difficulties in synthesizing semiconducting MXenes in aqueous solutions. The substitution of even minuscule amounts of terminal groups for oxygen in Sc-based MXenes and F/OH terminations in Ti-based MXene renders these materials conductive.

While for Ti$_{2}$C we found only one surface composition with a nonzero band gap, semiconducting Sc-based MXenes can have a variety of patterns and compositions of F and OH groups. We present the effect of these concentrations and patterns in Fig.~\ref{fig:Sc_gap_con}, which shows the band gap of Sc$_2$CF$_{2c}$(OH)$_{2(1-c)}$ as a function 
of F-group surface coverage $c$ for two types of coverages. Fig.~\ref{fig:Sc_gap_con} a) represents a gradual substitution of OH-groups by F mirrored on both sides of the MXene while on the other hand, Fig.~\ref{fig:Sc_gap_con} b) shows the case when the substitution first occurs on the side until saturation and then the OH groups on the other side are gradually substituted by F-terminations. These two substitution paths demonstrate the behavior of Sc$_2$C when: a) both sides of the MXene are exposed to the same chemical environment e.g. free floating in solution and b) when only one side is exposed to changes, e.g., Janus structure preparation. To ensure consistency with methodology in our previous studies~\cite{Sakhraoui2022}, for the calculation of band structures of semiconducting MXenes we have used spin-polarized DFT and found all structures to have a non-magnetic ground state.

When the substitution of OH groups for F-terminations is mirrored (Fig.~\ref{fig:Sc_gap_con} a)) the first substitution (Sc$_2$CF$_{0.22}$(OH)$_{1.78}$) shifts the band gap from pure OH-terminated Sc$_2$C, which is a direct material with the band gap in the $\Gamma$ point ($E_{g}^{d}$ = 0.54~eV), to 0.88~eV while maintaining the the direct band gap character. Subsequent substitutions change the band structure to resemble that of Sc$_2$CF$_{2}$ and the minimal gap becomes indirect ($E_{g}^{i}$=1.04~eV from $\Gamma$ to M) and both the indirect and direct band gaps ($E_{g}^{d}$ = 1.21~eV in the M point), linearly increase to values of Sc$_2$CF$_2$ ($E_{g}^{d}$ = 1.53~eV,  $E_{g}^{i}$ = 1.09~eV). In the case of Fig.~\ref{fig:Sc_gap_con} b), structures with single side F-group terminations exhibit band structures and gaps almost identical to that of Sc$_2$C(OH)$_{2}$. The gap linearly increases with the increasing number of F-terminations until a composition of Sc$_2$CF(OH) is achieved. After that, the gradual OH/F substitution on the other side causes the gaps and the material's behavior is identical to Fig.~\ref{fig:Sc_gap_con} a). This shows that for Sc$_2$CF$_{x}$(OH)$_{2-x}$ to become an indirect band gap semiconductor, the presence of F-terminations on both sides is required.

The band structures illustrating this transition are shown in Fig.~\ref{fig:FOH_bs} and in more detail in Fig.~S6. The change in band structure character is facilitated by valence bands formed from form oxygen orbitals around the gamma point. These bands gradually descend with the increasing number of OH-terminal groups until they become lower than the previously lowest valence band in the M-point.

An interesting feature of the presented gaps is that the variation in band gaps of structures with identical composition but varying patterns was on average 0.01 eV (Tab.~S5 and Tab.~S6), showing that the terminal group patterns do not play a role in determining the band gap. The main factor that governs the gap in Sc$_2$C is the surface concentration of specific groups.

\section{Conclusions}
In this study, we have sought to explain and quantify the influence of terminal groups on MXene chemistry. The first conclusion derived from our work pertains to the bond strength order and the stabilizing impact a terminal group exerts on the structure, i.e., the bond dissociation energy ($BDE$). The three most common terminal groups (T = O, F, and OH) provided the order for Ti-based and V-based MXenes to be  $BDE_{O} > BDE_{F} > BDE_{OH}$, while for Sc-based MXenes we found the general order to be $BDE_{F} > BDE_{O} > BDE_{OH}$ (and for conductive compositions order of $BDE_{O} > BDE_{F} > BDE_{OH}$). The arrangement for Ti-based MXenes aligns well with prior research and highlights a distinct energetic preference against the hydroxyl termination in all three metal-based MXenes. The terminal groups exhibit relatively strong bonds with the MXene (3-8~eV), prompting a need for further exploration into their nature.

In addition to these findings, another significant conclusion is that the terminal groups do not influence each other through the carbon layer. Bader charge analysis and structural analysis show that the terminal groups have a profound effect on the oxidation state and the carbon-metal distances on their side but the opposite side remains unaffected. Based on the reported behavior, terminal groups in higher-order MXenes should be affected to an even lesser degree by compositional variation on the other side of the MXene.

This fact allowed us to significantly reduce the structural space for 3$\times$3 supercell models when searching for an equilibrium surface group composition. The calculations of cohesion energy of mixed surface MXenes demonstrated that the interaction between the groups within the layer is negligible compared to their binding energies. Thus the ternary graphs obtained for all three (Sc-, V-, and Ti-) MXenes have smooth transitions between various T-group concentrations and do not depend on the surface patterns. While other external factors may affect pattern formation in experimental samples, there is no energetic reason for pattern formation on MXene surfaces which is supported by other theoretical~\cite{Ibragimova2019,Ibragimova2021} and experimental~\cite{Wang2015,Hope2016} studies. The cohesion energy of an MXenes is dominantly defined by its composition. Ti- and V-based MXenes prefer surface compositions with high concentrations of O-groups while the equilibrium composition for Sc was determined to be Sc$_{2}$CF$_{0.54}$O$_{1.46}$. Pattern variations have a negligible impact on electronic structure and cohesion energy, ergo, the same composition but varying patterns result in almost identical band gaps and cohesion energies. On the other hand, the properties of the material as a whole, e.g., the band gap, are significantly affected by the changing surface group composition.

A question that has not been addressed by previous studies is the effect of mixed compositions on the band gap of semi-conducting MXenes. While it is generally known that Sc-based MXenes exhibit semi-conducting behavior when terminated by F and OH groups and Ti-MXenes when covered by oxygen, the opening and closing of the band gap based on the surface composition has not been investigated anywhere else. We found that the band gap is extremely sensitive to the presence of conduction-inducing terminal groups. For Ti$_{2}$C surface concentrations of OH/F as low as 1\% (7$\times$7 supercell with one oxygen group substituted with F or OH) lead to the closing of the band gap. Similarly, for Sc$_{2}$C, any presence of even one O-terminal group in a 7$\times$7 supercell resulted in the structure being conductive. This may explain the experimental difficulties in obtaining semiconducting MXenes as even slight chemical changes in their surface composition result in a conductive MXene. If someone should succeed in obtaining an Sc$_{2}$C with only F and OH terminations, either through highly controlled CVD or specific surface treatments, the resulting band structure and gap of the material will strongly depend on the presence of these groups on either side of the structure.

Sc$_2$CF$_{x}$(OH)$_{2-x}$ MXenes exhibit an indirect band gap (direct band gap $\approx$ 1.5 eV and a smaller indirect band gap $\approx$ 1 eV ) if there are fluorine groups present on both sides of the MXene as shown in Fig.~\ref{fig:Sc_gap_con}. Sc$_2$CF$_{x}$(OH)$_{2-x}$ are direct band gap semiconductors only when F-terminal groups are confined to one side. The size of the band gap is governed by the induction effect of the F-terminations as a linear function of F surface concentration but only in the case of the indirect materials.\\

The presented results show an intriguing feature of MXenes in that even the thinnest MXenes can have two chemically independent sides that can be freely modified, and experience different chemical environments without affecting each other. Another feature that was uncovered by this work is the high sensitivity of semiconducting MXenes to certain terminal groups which may be conductive in their future synthesis and possible utilization as chemical sensors.


\section*{Conflicts of interest}
There are no conflicts to declare.

\section{Acknowledgments}
This work was supported by the Czech Science Foundation (21-28709S). The calculations were performed at IT4Innovations National Supercomputing Center (e-INFRA CZ, ID:90140).



\balance


\bibliography{rsc} 

\providecommand*{\mcitethebibliography}{\thebibliography}
\csname @ifundefined\endcsname{endmcitethebibliography}
{\let\endmcitethebibliography\endthebibliography}{}
\begin{mcitethebibliography}{51}
\providecommand*{\natexlab}[1]{#1}
\providecommand*{\mciteSetBstSublistMode}[1]{}
\providecommand*{\mciteSetBstMaxWidthForm}[2]{}
\providecommand*{\mciteBstWouldAddEndPuncttrue}
  {\def\EndOfBibitem{\unskip.}}
\providecommand*{\mciteBstWouldAddEndPunctfalse}
  {\let\EndOfBibitem\relax}
\providecommand*{\mciteSetBstMidEndSepPunct}[3]{}
\providecommand*{\mciteSetBstSublistLabelBeginEnd}[3]{}
\providecommand*{\EndOfBibitem}{}
\mciteSetBstSublistMode{f}
\mciteSetBstMaxWidthForm{subitem}
{(\emph{\alph{mcitesubitemcount}})}
\mciteSetBstSublistLabelBeginEnd{\mcitemaxwidthsubitemform\space}
{\relax}{\relax}

\bibitem[Anasori and Gogotsi(2022)]{Anasori2022}
B.~Anasori and Y.~Gogotsi, MXenes: trends, growth, and future directions,
  \emph{Graphene and 2D Materials}, 2022, \textbf{7}, 75--79\relax
\mciteBstWouldAddEndPuncttrue
\mciteSetBstMidEndSepPunct{\mcitedefaultmidpunct}
{\mcitedefaultendpunct}{\mcitedefaultseppunct}\relax
\EndOfBibitem
\bibitem[De \emph{et~al.}(2022)De, Acharya, Sahoo, Shim, and
  Nayak]{Shrabani2022}
S.~De, S.~Acharya, S.~Sahoo, J.-J. Shim and G.~C. Nayak, From 0D to 3D MXenes:
  their diverse syntheses{,} morphologies and applications, \emph{Materials
  Chemistry Frontiers}, 2022, \textbf{6}, 818--842\relax
\mciteBstWouldAddEndPuncttrue
\mciteSetBstMidEndSepPunct{\mcitedefaultmidpunct}
{\mcitedefaultendpunct}{\mcitedefaultseppunct}\relax
\EndOfBibitem
\bibitem[Khanal and Irle(2023)]{Khanal2023}
R.~Khanal and S.~Irle, Effect of surface functional groups on MXene
  conductivity, \emph{The Journal of Chemical Physics}, 2023, \textbf{158},
  194701\relax
\mciteBstWouldAddEndPuncttrue
\mciteSetBstMidEndSepPunct{\mcitedefaultmidpunct}
{\mcitedefaultendpunct}{\mcitedefaultseppunct}\relax
\EndOfBibitem
\bibitem[Brette \emph{et~al.}(2023)Brette, Kourati, Paris, Loupias,
  C\'{e}l\'{e}rier, Cabioc'h, Deschamps, Boucher, and Mauchamp]{Brette2023}
F.~Brette, D.~Kourati, M.~Paris, L.~Loupias, S.~C\'{e}l\'{e}rier, T.~Cabioc'h,
  M.~Deschamps, F.~Boucher and V.~Mauchamp, Assessing the Surface Chemistry of
  2D Transition Metal Carbides (MXenes): A Combined Experimental/Theoretical
  13C Solid State NMR Approach, \emph{Journal of the American Chemical
  Society}, 2023, \textbf{145}, 4003–4014\relax
\mciteBstWouldAddEndPuncttrue
\mciteSetBstMidEndSepPunct{\mcitedefaultmidpunct}
{\mcitedefaultendpunct}{\mcitedefaultseppunct}\relax
\EndOfBibitem
\bibitem[Wang \emph{et~al.}(2021)Wang, Ong, Naguib, and Wu]{Wang2021}
X.~Wang, G.~M. Ong, M.~Naguib and J.~Wu, Theoretical Insights into MXene
  Termination and Surface Charge Regulation, \emph{The Journal of Physical
  Chemistry C}, 2021, \textbf{125}, 21771–21779\relax
\mciteBstWouldAddEndPuncttrue
\mciteSetBstMidEndSepPunct{\mcitedefaultmidpunct}
{\mcitedefaultendpunct}{\mcitedefaultseppunct}\relax
\EndOfBibitem
\bibitem[Tang \emph{et~al.}(2021)Tang, Wang, and Zhang]{Tang2021}
C.~Tang, X.~Wang and S.~Zhang, Research on metallic chalcogen-functionalized
  monolayer-puckered V2CX2 (X = S, Se, and Te) as promising Li-ion battery
  anode materials, \emph{Materials Chemistry Frontiers}, 2021, \textbf{5},
  4672–4681\relax
\mciteBstWouldAddEndPuncttrue
\mciteSetBstMidEndSepPunct{\mcitedefaultmidpunct}
{\mcitedefaultendpunct}{\mcitedefaultseppunct}\relax
\EndOfBibitem
\bibitem[Ketolainen and Karlick\'{y}(2022)]{Ketolainen2022}
T.~Ketolainen and F.~Karlick\'{y}, Optical gaps and excitons in semiconducting
  transition metal carbides (MXenes), \emph{Journal of Materials Chemistry C},
  2022, \textbf{10}, 3919–3928\relax
\mciteBstWouldAddEndPuncttrue
\mciteSetBstMidEndSepPunct{\mcitedefaultmidpunct}
{\mcitedefaultendpunct}{\mcitedefaultseppunct}\relax
\EndOfBibitem
\bibitem[Khazaei \emph{et~al.}(2017)Khazaei, Ranjbar, Arai, Sasaki, and
  Yunoki]{Khazaei2017}
M.~Khazaei, A.~Ranjbar, M.~Arai, T.~Sasaki and S.~Yunoki, Electronic properties
  and applications of MXenes: a theoretical review, \emph{Journal of Materials
  Chemistry C}, 2017, \textbf{5}, 2488–2503\relax
\mciteBstWouldAddEndPuncttrue
\mciteSetBstMidEndSepPunct{\mcitedefaultmidpunct}
{\mcitedefaultendpunct}{\mcitedefaultseppunct}\relax
\EndOfBibitem
\bibitem[Hart \emph{et~al.}(2019)Hart, Hantanasirisakul, Lang, Anasori, Pinto,
  Pivak, van Omme, May, Gogotsi, and Taheri]{Hart2019}
J.~L. Hart, K.~Hantanasirisakul, A.~C. Lang, B.~Anasori, D.~Pinto, Y.~Pivak,
  J.~T. van Omme, S.~J. May, Y.~Gogotsi and M.~L. Taheri, Control of MXenes'
  electronic properties through termination and intercalation, \emph{Nature
  Communications}, 2019, \textbf{10}, 522\relax
\mciteBstWouldAddEndPuncttrue
\mciteSetBstMidEndSepPunct{\mcitedefaultmidpunct}
{\mcitedefaultendpunct}{\mcitedefaultseppunct}\relax
\EndOfBibitem
\bibitem[Handoko \emph{et~al.}(2020)Handoko, Chen, Lum, Zhang, Anasori, and
  Seh]{Handoko2020}
A.~D. Handoko, H.~Chen, Y.~Lum, Q.~Zhang, B.~Anasori and Z.~W. Seh,
  Two-Dimensional Titanium and Molybdenum Carbide MXenes as Electrocatalysts
  for CO2 Reduction, \emph{iScience}, 2020, \textbf{23}, 101181\relax
\mciteBstWouldAddEndPuncttrue
\mciteSetBstMidEndSepPunct{\mcitedefaultmidpunct}
{\mcitedefaultendpunct}{\mcitedefaultseppunct}\relax
\EndOfBibitem
\bibitem[Urbankowski \emph{et~al.}(2016)Urbankowski, Anasori, Makaryan, Er,
  Kota, Walsh, Zhao, Shenoy, Barsoum, and Gogotsi]{Urbankowski2016}
P.~Urbankowski, B.~Anasori, T.~Makaryan, D.~Er, S.~Kota, P.~L. Walsh, M.~Zhao,
  V.~B. Shenoy, M.~W. Barsoum and Y.~Gogotsi, Synthesis of two-dimensional
  titanium nitride Ti4N3(MXene), \emph{Nanoscale}, 2016, \textbf{8},
  11385–11391\relax
\mciteBstWouldAddEndPuncttrue
\mciteSetBstMidEndSepPunct{\mcitedefaultmidpunct}
{\mcitedefaultendpunct}{\mcitedefaultseppunct}\relax
\EndOfBibitem
\bibitem[Halim \emph{et~al.}(2016)Halim, Cook, Naguib, Eklund, Gogotsi, Rosen,
  and Barsoum]{Halim2016}
J.~Halim, K.~M. Cook, M.~Naguib, P.~Eklund, Y.~Gogotsi, J.~Rosen and M.~W.
  Barsoum, X-ray photoelectron spectroscopy of select multi-layered transition
  metal carbides (MXenes), \emph{Applied Surface Science}, 2016, \textbf{362},
  406–417\relax
\mciteBstWouldAddEndPuncttrue
\mciteSetBstMidEndSepPunct{\mcitedefaultmidpunct}
{\mcitedefaultendpunct}{\mcitedefaultseppunct}\relax
\EndOfBibitem
\bibitem[Hope \emph{et~al.}(2016)Hope, Forse, Griffith, Lukatskaya, Ghidiu,
  Gogotsi, and Grey]{Hope2016}
M.~A. Hope, A.~C. Forse, K.~J. Griffith, M.~R. Lukatskaya, M.~Ghidiu,
  Y.~Gogotsi and C.~P. Grey, NMR reveals the surface functionalisation of
  Ti3C2MXene, \emph{Physical Chemistry Chemical Physics}, 2016, \textbf{18},
  5099–5102\relax
\mciteBstWouldAddEndPuncttrue
\mciteSetBstMidEndSepPunct{\mcitedefaultmidpunct}
{\mcitedefaultendpunct}{\mcitedefaultseppunct}\relax
\EndOfBibitem
\bibitem[Hu \emph{et~al.}(2018)Hu, Hu, Gao, Li, and Wang]{Hu2018}
T.~Hu, M.~Hu, B.~Gao, W.~Li and X.~Wang, Screening Surface Structure of MXenes
  by High-Throughput Computation and Vibrational Spectroscopic Confirmation,
  \emph{The Journal of Physical Chemistry C}, 2018, \textbf{122},
  18501–18509\relax
\mciteBstWouldAddEndPuncttrue
\mciteSetBstMidEndSepPunct{\mcitedefaultmidpunct}
{\mcitedefaultendpunct}{\mcitedefaultseppunct}\relax
\EndOfBibitem
\bibitem[Ibragimova \emph{et~al.}(2021)Ibragimova, Erhart, Rinke, and
  Komsa]{Ibragimova2021}
R.~Ibragimova, P.~Erhart, P.~Rinke and H.-P. Komsa, Surface Functionalization
  of 2D MXenes: Trends in Distribution, Composition, and Electronic Properties,
  \emph{The Journal of Physical Chemistry Letters}, 2021, \textbf{12},
  2377–2384\relax
\mciteBstWouldAddEndPuncttrue
\mciteSetBstMidEndSepPunct{\mcitedefaultmidpunct}
{\mcitedefaultendpunct}{\mcitedefaultseppunct}\relax
\EndOfBibitem
\bibitem[Li \emph{et~al.}(2021)Li, Guo, Wang, Ma, and Wang]{Li2021}
C.~Li, J.~Guo, C.~Wang, D.~Ma and B.~Wang, Two-dimensional Janus Ti3C2F O2-
  MXene with tunable electronic and mechanical properties, \emph{Scripta
  Materialia}, 2021, \textbf{194}, 113688\relax
\mciteBstWouldAddEndPuncttrue
\mciteSetBstMidEndSepPunct{\mcitedefaultmidpunct}
{\mcitedefaultendpunct}{\mcitedefaultseppunct}\relax
\EndOfBibitem
\bibitem[Jin \emph{et~al.}(2018)Jin, Wu, and Wang]{Jin2018}
W.~Jin, S.~Wu and Z.~Wang, Structural, electronic and mechanical properties of
  two-dimensional Janus transition metal carbides and nitrides, \emph{Physica
  E: Low-dimensional Systems and Nanostructures}, 2018, \textbf{103},
  307–313\relax
\mciteBstWouldAddEndPuncttrue
\mciteSetBstMidEndSepPunct{\mcitedefaultmidpunct}
{\mcitedefaultendpunct}{\mcitedefaultseppunct}\relax
\EndOfBibitem
\bibitem[Frey \emph{et~al.}(2019)Frey, Bandyopadhyay, Kumar, Anasori, Gogotsi,
  and Shenoy]{Frey2019}
N.~C. Frey, A.~Bandyopadhyay, H.~Kumar, B.~Anasori, Y.~Gogotsi and V.~B.
  Shenoy, Surface-Engineered MXenes: Electric Field Control of Magnetism and
  Enhanced Magnetic Anisotropy, \emph{ACS Nano}, 2019, \textbf{13},
  2831–2839\relax
\mciteBstWouldAddEndPuncttrue
\mciteSetBstMidEndSepPunct{\mcitedefaultmidpunct}
{\mcitedefaultendpunct}{\mcitedefaultseppunct}\relax
\EndOfBibitem
\bibitem[Caffrey(2018)]{Caffrey2018}
N.~M. Caffrey, Effect of mixed surface terminations on the structural and
  electrochemical properties of two-dimensional Ti3C2T2 and V2CT2 MXenes
  multilayers, \emph{Nanoscale}, 2018, \textbf{10}, 13520–13530\relax
\mciteBstWouldAddEndPuncttrue
\mciteSetBstMidEndSepPunct{\mcitedefaultmidpunct}
{\mcitedefaultendpunct}{\mcitedefaultseppunct}\relax
\EndOfBibitem
\bibitem[Yadav \emph{et~al.}(2022)Yadav, Vikram, Singh, and Alam]{Yadav2022}
A.~Yadav, Vikram, N.~Singh and A.~Alam, Mixed-Functionalized Sc2CTx (T=O, OH,
  F) MXene for Electrocatalytic CO2 Reduction: Insight from First-Principles
  Calculations, \emph{Phys. Rev. Appl.}, 2022, \textbf{18}, 024020\relax
\mciteBstWouldAddEndPuncttrue
\mciteSetBstMidEndSepPunct{\mcitedefaultmidpunct}
{\mcitedefaultendpunct}{\mcitedefaultseppunct}\relax
\EndOfBibitem
\bibitem[Cui \emph{et~al.}(2024)Cui, Yu, Li, Zhang, and Cui]{Cui2024}
C.-C. Cui, S.-F. Yu, X.-H. Li, R.-Z. Zhang and H.-L. Cui, Effect of mixed
  surface terminations on the work function and quantum capacitance of Sc2CT2
  monolayer, \emph{Surface Science}, 2024, \textbf{740}, 122413\relax
\mciteBstWouldAddEndPuncttrue
\mciteSetBstMidEndSepPunct{\mcitedefaultmidpunct}
{\mcitedefaultendpunct}{\mcitedefaultseppunct}\relax
\EndOfBibitem
\bibitem[Cui \emph{et~al.}(2023)Cui, Li, Zhang, Cui, and Yan]{Cui2023}
X.-H. Cui, X.-H. Li, R.-Z. Zhang, H.-L. Cui and H.-T. Yan, Theoretical insight
  into the electronic, optical, and photocatalytic properties and quantum
  capacitance of Sc2CT2 (T = F, P, Cl, Se, Br, O, Si, S, OH) MXenes,
  \emph{Vacuum}, 2023, \textbf{207}, 111615\relax
\mciteBstWouldAddEndPuncttrue
\mciteSetBstMidEndSepPunct{\mcitedefaultmidpunct}
{\mcitedefaultendpunct}{\mcitedefaultseppunct}\relax
\EndOfBibitem
\bibitem[Kumar and Schwingenschl\"ogl(2016)]{Kumar2016}
S.~Kumar and U.~Schwingenschl\"ogl, Thermoelectric performance of
  functionalized Sc2C MXenes, \emph{Phys. Rev. B}, 2016, \textbf{94},
  035405\relax
\mciteBstWouldAddEndPuncttrue
\mciteSetBstMidEndSepPunct{\mcitedefaultmidpunct}
{\mcitedefaultendpunct}{\mcitedefaultseppunct}\relax
\EndOfBibitem
\bibitem[Sakhraoui and Karlick\'{y}(2024)]{Sakhraoui2024}
T.~Sakhraoui and F.~Karlick\'{y}, Prediction of induced magnetism in 2D Ti2C
  based MXenes by manipulating the mixed surface functionalization and metal
  substitution computed by xTB model Hamiltonian of the DFTB method,
  \emph{Physical Chemistry Chemical Physics}, 2024, \textbf{26},
  12862–12868\relax
\mciteBstWouldAddEndPuncttrue
\mciteSetBstMidEndSepPunct{\mcitedefaultmidpunct}
{\mcitedefaultendpunct}{\mcitedefaultseppunct}\relax
\EndOfBibitem
\bibitem[Kresse and Hafner(1993)]{Kresse_PhysRevB_47_1993}
G.~Kresse and J.~Hafner, Ab initio Molecular Dynamics For Liquid Metals.,
  \emph{Physical Reviews B}, 1993, \textbf{47}, 558\relax
\mciteBstWouldAddEndPuncttrue
\mciteSetBstMidEndSepPunct{\mcitedefaultmidpunct}
{\mcitedefaultendpunct}{\mcitedefaultseppunct}\relax
\EndOfBibitem
\bibitem[Kresse and Hafner(1994)]{Kresse_PhysRevB_49_1994}
G.~Kresse and J.~Hafner, Ab initio Molecular-Dynamics Simulation of the
  Liquid-Metal-Amorphous-Semiconductor Transition in Germanium., \emph{Physical
  Reviews B}, 1994, \textbf{49}, 14251\relax
\mciteBstWouldAddEndPuncttrue
\mciteSetBstMidEndSepPunct{\mcitedefaultmidpunct}
{\mcitedefaultendpunct}{\mcitedefaultseppunct}\relax
\EndOfBibitem
\bibitem[Kresse and Furthm\"{u}ller(1996)]{Kresse_Comp_Mat_Sci_6_1996}
G.~Kresse and J.~Furthm\"{u}ller, Efficiency of Ab-initio Total Energy
  Calculations for Metals and Semiconductors Using a Plane-Wave Basis Set,
  \emph{Computational Material Science}, 1996, \textbf{6}, 15\relax
\mciteBstWouldAddEndPuncttrue
\mciteSetBstMidEndSepPunct{\mcitedefaultmidpunct}
{\mcitedefaultendpunct}{\mcitedefaultseppunct}\relax
\EndOfBibitem
\bibitem[Kresse and Furthm\"{u}ller(1996)]{Kresse_PhysRevB_54_1996}
G.~Kresse and J.~Furthm\"{u}ller, Efficient Iterative Schemes For Ab initio
  total-Energy Calculations Using A Plane-Wave Basis Set., \emph{Physical
  Reviews B}, 1996, \textbf{54}, 11169\relax
\mciteBstWouldAddEndPuncttrue
\mciteSetBstMidEndSepPunct{\mcitedefaultmidpunct}
{\mcitedefaultendpunct}{\mcitedefaultseppunct}\relax
\EndOfBibitem
\bibitem[Perdew \emph{et~al.}(1996)Perdew, Burke, and
  Ernzerhof]{Perdew_PhysRewLet_77_1996}
J.~P. Perdew, K.~Burke and M.~Ernzerhof, Generalized Gradient Approximation
  Made Simple., \emph{Physical Review Letters}, 1996, \textbf{77}, 3865\relax
\mciteBstWouldAddEndPuncttrue
\mciteSetBstMidEndSepPunct{\mcitedefaultmidpunct}
{\mcitedefaultendpunct}{\mcitedefaultseppunct}\relax
\EndOfBibitem
\bibitem[Perdew \emph{et~al.}(1999)Perdew, Burke, and
  Ernzerhof]{Perdew_PhysRewLet_78_1997}
J.~P. Perdew, K.~Burke and M.~Ernzerhof, Erratum: Generalized Gradient
  Approximation Made Simple., \emph{Physical Review Letters}, 1999,
  \textbf{78}, 1396\relax
\mciteBstWouldAddEndPuncttrue
\mciteSetBstMidEndSepPunct{\mcitedefaultmidpunct}
{\mcitedefaultendpunct}{\mcitedefaultseppunct}\relax
\EndOfBibitem
\bibitem[Sun \emph{et~al.}(2015)Sun, Ruzsinszky, and Perdew]{SCAN2}
J.~Sun, A.~Ruzsinszky and J.~Perdew, Strongly Constrained and Appropriately
  Normed Semilocal Density Functional, \emph{Physical Review Letters}, 2015,
  \textbf{115}, 036402\relax
\mciteBstWouldAddEndPuncttrue
\mciteSetBstMidEndSepPunct{\mcitedefaultmidpunct}
{\mcitedefaultendpunct}{\mcitedefaultseppunct}\relax
\EndOfBibitem
\bibitem[Blochl(1994)]{Blochl_PhysRevB_50_1994}
P.~E. Blochl, Projector Augmented-Wave Method., \emph{Physical Reviews B},
  1994, \textbf{50}, 1795\relax
\mciteBstWouldAddEndPuncttrue
\mciteSetBstMidEndSepPunct{\mcitedefaultmidpunct}
{\mcitedefaultendpunct}{\mcitedefaultseppunct}\relax
\EndOfBibitem
\bibitem[Kresse and Joubert(1999)]{Kresse_PhysRevB_59_1999}
G.~Kresse and D.~Joubert, From Ultrasoft Pseudopotentials to the Projector
  Augmented-Wave Method., \emph{Physical Reviews B}, 1999, \textbf{59},
  1758\relax
\mciteBstWouldAddEndPuncttrue
\mciteSetBstMidEndSepPunct{\mcitedefaultmidpunct}
{\mcitedefaultendpunct}{\mcitedefaultseppunct}\relax
\EndOfBibitem
\bibitem[Pulay(1980)]{Pulay1980}
P.~Pulay, Convergence acceleration of iterative sequences. the case of scf
  iteration, \emph{Chemical Physics Letters}, 1980, \textbf{73},
  393–398\relax
\mciteBstWouldAddEndPuncttrue
\mciteSetBstMidEndSepPunct{\mcitedefaultmidpunct}
{\mcitedefaultendpunct}{\mcitedefaultseppunct}\relax
\EndOfBibitem
\bibitem[Dubeck\'{y} \emph{et~al.}(2023)Dubeck\'{y}, Min\'{a}rik, and
  Karlick\'{y}]{Dubeck2023}
M.~Dubeck\'{y}, S.~Min\'{a}rik and F.~Karlick\'{y}, Benchmarking fundamental
  gap of Sc2C(OH)2 MXene by many-body methods, \emph{The Journal of Chemical
  Physics}, 2023, \textbf{158}, 054703\relax
\mciteBstWouldAddEndPuncttrue
\mciteSetBstMidEndSepPunct{\mcitedefaultmidpunct}
{\mcitedefaultendpunct}{\mcitedefaultseppunct}\relax
\EndOfBibitem
\bibitem[Zhang \emph{et~al.}(2021)Zhang, Sa, Miao, Zhou, and Sun]{Zhang2021}
Y.~Zhang, B.~Sa, N.~Miao, J.~Zhou and Z.~Sun, Computational mining of Janus
  Sc2C-based MXenes for spintronic, photocatalytic, and solar cell
  applications, \emph{Journal of Materials Chemistry A}, 2021, \textbf{9},
  10882–10892\relax
\mciteBstWouldAddEndPuncttrue
\mciteSetBstMidEndSepPunct{\mcitedefaultmidpunct}
{\mcitedefaultendpunct}{\mcitedefaultseppunct}\relax
\EndOfBibitem
\bibitem[Kumar \emph{et~al.}(2024)Kumar, Kolos, Bhattacharya, and
  Karlick\'{y}]{Kumar2024}
N.~Kumar, M.~Kolos, S.~Bhattacharya and F.~Karlick\'{y}, Excitons, Optical
  Spectra, and Electronic Properties of Semiconducting Hf-based MXenes,
  \emph{The Journal of Chemical Physics}, 2024, \textbf{160}, 124707\relax
\mciteBstWouldAddEndPuncttrue
\mciteSetBstMidEndSepPunct{\mcitedefaultmidpunct}
{\mcitedefaultendpunct}{\mcitedefaultseppunct}\relax
\EndOfBibitem
\bibitem[Khazaei \emph{et~al.}(2012)Khazaei, Arai, Sasaki, Chung,
  Venkataramanan, Estili, Sakka, and Kawazoe]{Khazaei2012}
M.~Khazaei, M.~Arai, T.~Sasaki, C.-Y. Chung, N.~S. Venkataramanan, M.~Estili,
  Y.~Sakka and Y.~Kawazoe, Novel Electronic and Magnetic Properties of
  Two-Dimensional Transition Metal Carbides and Nitrides, \emph{Advanced
  Functional Materials}, 2012, \textbf{23}, 2185–2192\relax
\mciteBstWouldAddEndPuncttrue
\mciteSetBstMidEndSepPunct{\mcitedefaultmidpunct}
{\mcitedefaultendpunct}{\mcitedefaultseppunct}\relax
\EndOfBibitem
\bibitem[Sakhraoui and Karlick\'{y}(2022)]{Sakhraoui2022}
T.~Sakhraoui and F.~Karlick\'{y}, Electronic Nature Transition and Magnetism
  Creation in Vacancy-Defected Ti2CO2 MXene under Biaxial Strain: A DFTB + U
  Study, \emph{ACS Omega}, 2022, \textbf{7}, 42221–42232\relax
\mciteBstWouldAddEndPuncttrue
\mciteSetBstMidEndSepPunct{\mcitedefaultmidpunct}
{\mcitedefaultendpunct}{\mcitedefaultseppunct}\relax
\EndOfBibitem
\bibitem[Tang \emph{et~al.}(2009)Tang, Sanville, and Henkelman]{bader}
W.~Tang, E.~Sanville and G.~Henkelman, A grid-based Bader analysis algorithm
  without lattice bias, \emph{Journal of Physics: Condensed Matter}, 2009,
  \textbf{21}, 084204\relax
\mciteBstWouldAddEndPuncttrue
\mciteSetBstMidEndSepPunct{\mcitedefaultmidpunct}
{\mcitedefaultendpunct}{\mcitedefaultseppunct}\relax
\EndOfBibitem
\bibitem[Ban()]{BandUnfol}
\url{https://github.com/QijingZheng/VaspBandUnfolding}\relax
\mciteBstWouldAddEndPuncttrue
\mciteSetBstMidEndSepPunct{\mcitedefaultmidpunct}
{\mcitedefaultendpunct}{\mcitedefaultseppunct}\relax
\EndOfBibitem
\bibitem[Popescu and Zunger(2012)]{Popescu2012}
V.~Popescu and A.~Zunger, Extracting E versus k effective band structure from
  supercell calculations on alloys and impurities, \emph{Physical Review B},
  2012, \textbf{85}, 085201\relax
\mciteBstWouldAddEndPuncttrue
\mciteSetBstMidEndSepPunct{\mcitedefaultmidpunct}
{\mcitedefaultendpunct}{\mcitedefaultseppunct}\relax
\EndOfBibitem
\bibitem[Yusupov \emph{et~al.}(2023)Yusupov, Bj\"{o}rk, and Rosen]{Yusupov2023}
K.~Yusupov, J.~Bj\"{o}rk and J.~Rosen, A systematic study of work function and
  electronic properties of MXenes from first principles, \emph{Nanoscale
  Advances}, 2023, \textbf{5}, 3976–3984\relax
\mciteBstWouldAddEndPuncttrue
\mciteSetBstMidEndSepPunct{\mcitedefaultmidpunct}
{\mcitedefaultendpunct}{\mcitedefaultseppunct}\relax
\EndOfBibitem
\bibitem[Atkins \emph{et~al.}(2009)Atkins, Overton, Rourke, Weller, and
  Armstrong]{Atkins2009-pq}
P.~Atkins, T.~Overton, J.~Rourke, M.~Weller and F.~Armstrong, \emph{Shriver and
  Atkins' inorganic chemistry}, Oxford University Press, London, England, 5th
  edn, 2009\relax
\mciteBstWouldAddEndPuncttrue
\mciteSetBstMidEndSepPunct{\mcitedefaultmidpunct}
{\mcitedefaultendpunct}{\mcitedefaultseppunct}\relax
\EndOfBibitem
\bibitem[Wang \emph{et~al.}(2015)Wang, Naguib, Page, Wesolowski, and
  Gogotsi]{Wang2015}
H.-W. Wang, M.~Naguib, K.~Page, D.~J. Wesolowski and Y.~Gogotsi, Resolving the
  Structure of Ti3C2Tx MXenes through Multilevel Structural Modeling of the
  Atomic Pair Distribution Function, \emph{Chemistry of Materials}, 2015,
  \textbf{28}, 349–359\relax
\mciteBstWouldAddEndPuncttrue
\mciteSetBstMidEndSepPunct{\mcitedefaultmidpunct}
{\mcitedefaultendpunct}{\mcitedefaultseppunct}\relax
\EndOfBibitem
\bibitem[Fagerli \emph{et~al.}(2022)Fagerli, Wang, Grande, Kaland, Selbach,
  Wagner, and Wiik]{Fagerli2022}
F.~H. Fagerli, Z.~Wang, T.~Grande, H.~Kaland, S.~M. Selbach, N.~P. Wagner and
  K.~Wiik, Removing Fluoride-Terminations from Multilayered V2CTx MXene by Gas
  Hydrolyzation, \emph{ACS Omega}, 2022, \textbf{7}, 23790–23799\relax
\mciteBstWouldAddEndPuncttrue
\mciteSetBstMidEndSepPunct{\mcitedefaultmidpunct}
{\mcitedefaultendpunct}{\mcitedefaultseppunct}\relax
\EndOfBibitem
\bibitem[Champagne and Charlier(2020)]{Champagne2020}
A.~Champagne and J.-C. Charlier, Physical properties of 2D MXenes: from a
  theoretical perspective, \emph{Journal of Physics: Materials}, 2020,
  \textbf{3}, 032006\relax
\mciteBstWouldAddEndPuncttrue
\mciteSetBstMidEndSepPunct{\mcitedefaultmidpunct}
{\mcitedefaultendpunct}{\mcitedefaultseppunct}\relax
\EndOfBibitem
\bibitem[Zha \emph{et~al.}(2015)Zha, Luo, Li, Huang, He, Wen, and Du]{Zha2015}
X.-H. Zha, K.~Luo, Q.~Li, Q.~Huang, J.~He, X.~Wen and S.~Du, Role of the
  surface effect on the structural, electronic and mechanical properties of the
  carbide MXenes, \emph{EPL (Europhysics Letters)}, 2015, \textbf{111},
  26007\relax
\mciteBstWouldAddEndPuncttrue
\mciteSetBstMidEndSepPunct{\mcitedefaultmidpunct}
{\mcitedefaultendpunct}{\mcitedefaultseppunct}\relax
\EndOfBibitem
\bibitem[Kumar and Karlick\'{y}(2023)]{Kumar2023}
N.~Kumar and F.~Karlick\'{y}, Oxygen-terminated Ti3C2 MXene as an excitonic
  insulator, \emph{Applied Physics Letters}, 2023, \textbf{122}, 183102\relax
\mciteBstWouldAddEndPuncttrue
\mciteSetBstMidEndSepPunct{\mcitedefaultmidpunct}
{\mcitedefaultendpunct}{\mcitedefaultseppunct}\relax
\EndOfBibitem
\bibitem[Ibragimova \emph{et~al.}(2019)Ibragimova, Puska, and
  Komsa]{Ibragimova2019}
R.~Ibragimova, M.~J. Puska and H.-P. Komsa, pH-Dependent Distribution of
  Functional Groups on Titanium-Based MXenes, \emph{ACS Nano}, 2019,
  \textbf{13}, 9171–9181\relax
\mciteBstWouldAddEndPuncttrue
\mciteSetBstMidEndSepPunct{\mcitedefaultmidpunct}
{\mcitedefaultendpunct}{\mcitedefaultseppunct}\relax
\EndOfBibitem
\end{mcitethebibliography}
\bibliographystyle{rsc} 

\end{document}